\documentclass[aps,prx,reprint,twocolumn,floatfix]{revtex4-2}
\usepackage{graphicx}
\usepackage{dcolumn}
\usepackage{bm}
\usepackage[normalem]{ulem}
\usepackage{xcolor}
\usepackage{comment}
\usepackage{amsmath}
\usepackage{siunitx}
\usepackage{dsfont}
\usepackage{float}
\usepackage{placeins}
\usepackage[shortlabels]{enumitem}
\usepackage{soul}
\usepackage{booktabs}

\usepackage{hyperref}

\newcommand{\bluemark}[1] {\color{blue}#1\color{black}}

\DeclareSIUnit\angstrom{\text{Å}}

\usepackage{physics}

\begin{document}

\title{Imaging  spin  entanglement with  a scanning tunneling microscope}

\author{J. C. G. Henriques$^{1,2}$,  B. Viña-Baus\'a$^{3}$, I. Brihuega$^{3,4}$, J. Fern\'{a}ndez-Rossier$^1$\footnote{On permanent leave from Departamento de F\'{i}sica, Universidad de Alicante, 03690 San Vicente del Raspeig, Spain}$^,$\footnote{joaquin.fernandez-rossier@inl.int} }

\affiliation{$^1$International Iberian Nanotechnology Laboratory (INL), Av. Mestre Jos\'{e} Veiga, 4715-330 Braga, Portugal }
\affiliation{$^2$Universidade de Santiago de Compostela, 15782 Santiago de Compostela, Spain}

\affiliation{
$^3$ Departamento de F\'isica de la Materia Condensada,
Universidad Aut\'onoma de Madrid,
28049 Madrid, Spain
}
\affiliation{
$^4$ Condensed Matter Physics Center (IFIMAC) and
Instituto Nicol\'as Cabrera (INC),
Universidad Aut\'onoma de Madrid,
28049 Madrid, Spain
}

\date{\today}


\begin{abstract}
Artificial spin arrays assembled on surfaces enable atomic-scale studies of quantum correlations. Here we establish a hierarchy of connections between spin inelastic electron tunneling spectroscopy (IETS) and entanglement in spin-$1/2$ systems. We first show that the number of spectroscopically active transitions in a spin-$1/2$ dimer serves as an entanglement witness. Using the leading-order tunneling conductance for the tip placed on spin $i$, $dI(i)/dV$, we then relate spectral moments $\mathcal{M}^{(n)}(i)=\int V^n(d^2I(i)/dV^2)\,dV$ to quantum correlations. The zeroth moment probes the one-tangle in systems with a gapped, nondegenerate ground state, and gives information about its magnetization, without a spin-polarized tip. Comparison with experimental data of Hydrogen atoms on graphene shows good agreement with simulations. For rotationally invariant Heisenberg systems, the first moment yields thermal spin correlators $\langle\boldsymbol S(i)\cdot\boldsymbol S(j)\rangle$, enabling reconstruction of nearest-neighbor concurrence and spatial imaging of entanglement in spin chains. It also provides access to the internal energy and magnetic specific heat, both usable as entanglement witnesses. Applying this approach to experimental IETS data from nanographene spin chains, we find good agreement with theory. We further discuss how the second moment probes longer-range correlations and the inverse moment relates to local spin susceptibility. For spin-$1/2$ trimers, we connect inter-multiplet inelastic spectral weight to genuine tripartite entanglement quantified by the concurrence fill. Finally, we reconcile the local character of IETS with its sensitivity to nonlocal entanglement through an analogy with digital quantum circuits, where Hamiltonian evolution implements multiqubit operations. These results establish IETS as a tool for quantifying and imaging entanglement in engineered on-surface spin systems.
\end{abstract}
\maketitle

\section{Introduction}

Scanning tunneling microscopy (STM) was initially developed to image surfaces with atomic-scale resolution, exploiting the exponential dependence of the tunneling current on the tip--surface separation~\cite{binnig82}. Its spectroscopic capabilities subsequently provided access to the local electronic density of states~\cite{tersoff85}, including standing-wave patterns from which surface-state dispersions could be extracted~\cite{crommie93,hasegawa93}. STM-based inelastic electron tunneling spectroscopy (IETS) extended this capability to molecular vibrations~\cite{stipe98} and spin excitations of individual magnetic atoms~\cite{heinrich04,hirjibehedin07,khajetoorians11,bryant13}, engineered atomic arrays~\cite{hirjibehedin06,spinelli14,toskovic16,choi19,khajetoorians19}, and magnetic molecules~\cite{chen08,tsukahara09,huang25}. More recently, spin IETS has become a powerful probe of quantum magnetism in nanographenes, revealing exchange-coupled spins in individual structures and dimers~\cite{li19,mishra20,zheng20,zheng20b,song24,li26}, collective excitations in triangulene rings~\cite{Mishra21,hieulle21,li26b}, and fractional and topological excitations in molecular chains~\cite{Mishra21,zhao24,su25,yuan25,zhao25,fu25,vaxevani26} and ladders~\cite{zhao26}. IETS has also been used to probe the collective spin excitations in $S=1/2$ dimers and trimers made with chemisorbed atomic hydrogen in graphene\cite{vina26}.  Since spin IETS probes not only excitation energies but also the spin matrix elements connecting many-body eigenstates~\cite{fernandezrossier09}, it is natural to ask whether this information can be used to detect, quantify and, given the scanning nature of the experiment, image spin entanglement on-surfaces.

Quantum entanglement is ubiquitous in interacting quantum
many-body systems~\cite{amico08}.
The entangled ground states of certain spin Hamiltonians
also provide resources for quantum information processing:
AKLT states enable measurement-based quantum
computation~\cite{brennen08,wei11,miyake11}, the toric
code encodes quantum information protected against local
errors~\cite{kitaev03}, and the non-Abelian phase of the
Kitaev honeycomb model supports topological quantum gates
through anyon braiding~\cite{kitaev06}.
 Detecting and quantifying entanglement in solid-state systems has therefore attracted sustained interest~\cite{ghosh03,laurell25}. In quantum magnets, experimentally accessible entanglement witnesses include magnetic susceptibility~\cite{wiesniak05,brukner06}, energy, including its ground-state value~\cite{brukner04,dowling04,toth05}, and specific heat~\cite{wiesniak08}. Complementary approaches exploit spin correlations and structure factors accessible through neutron scattering~\cite{krammer09,cramer11,marty14}, that  has also been used to
quantify concurrence between molecular spin qubits
from their spectroscopic intensities~\cite{garlatti17}.  The connection between quantum Fisher information and dynamical spin susceptibility~\cite{hauke16} enables spectroscopic detection of multipartite entanglement, as demonstrated using inelastic neutron scattering~\cite{mathew20,scheie21,laurell21}. These developments establish conventional thermodynamic and spectroscopic probes as tools to access quantum entanglement in materials.

\begin{figure*}[t]
\centering
\includegraphics[width=\textwidth]{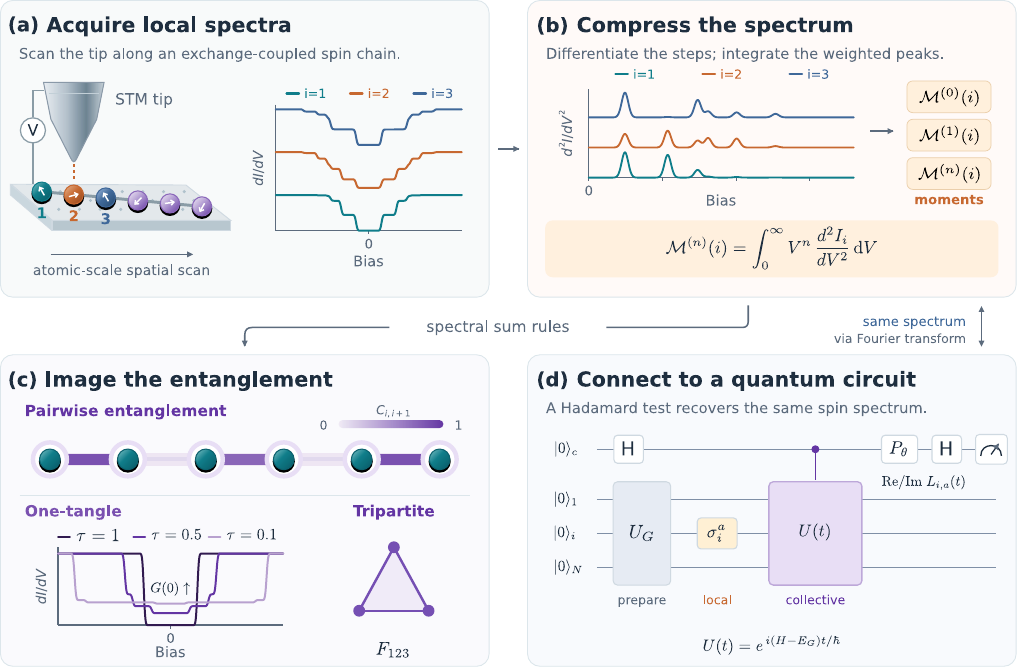}
\caption{
\textbf{From local tunnelling spectra to images of spin
entanglement.}
(a) STM-based IETS measured at different sites of an interacting spin-$1/2$ chain probes the same collective excitation energies with site-dependent intensities; spin arrows are schematic and do not represent static local magnetizations.
(b) Integration of the local $d^2I/dV^2$ spectra gives the spectral moments $\mathcal{M}^{(n)}(i)$. 
(c) The $\mathcal{M}^{(n)}(i)$ enable imaging of entanglement, for example through the one-tangle, $\tau_i$, and the bond concurrence $C_{i,j}$.
(d) A quantum circuit accesses the same spectral information through a local spin operation followed by controlled evolution under the interacting Hamiltonian, making explicit the many-body dynamics underlying the spectroscopic measurement. Spectra and entanglement maps are schematic.}
\label{fig:overview}
\end{figure*}
On-surface spins provide a uniquely tunable arena for engineering quantum spin models\cite{choi19,khajetoorians19,Mishra21,wang24,zhao24} with entangled states, while STM offers local spin-sensitive probes that naturally motivate the investigation of entanglement both at equilibrium and under coherent driving. Proposals using  electron spin resonance (ESR)-STM\cite{baumann15} address these complementary settings through susceptibility-based certification of equilibrium entanglement~\cite{delCastillo23} and protocols for generating and certifying entanglement in coherently driven spins~\cite{broekhoven24}. 

In this work we establish a framework to detect, quantify and image equilibrium spin entanglement using a much more widespread experimntal technique,  STM-based IETS with atomic-scale spatial resolution. Figure~\ref{fig:overview} summarizes our approach, connecting
local tunnelling spectra, spectral moments, and spatial maps
of spin entanglement.

The rest of the manuscript is organized as follows. 
In Sec. \ref{sec:entanglement} we briefly review the entanglement metrics that are relevant in this paper.
In Sec. \ref{sec:IETS} we  review the theory of IETS\cite{fernandezrossier09} that provides the background for this work.  In that section we already identify a very simple entanglement witness based on the number of inelastic excitation steps. 
In Sec. \ref{sec:M0} we introduce  the concept of $n$th-moment of the differential conductance, $\mathcal{M}^{(n)}$, and we discuss the relation between $\mathcal{M}^{(0)}$ and the one-tangle. 
In Sec. \ref{sec:M1} we demonstrate the relation between $\mathcal{M}^{(1)}$  and the spin correlators, concurrence  and energy-based witnesses.
In Sec. \ref{sec:experimental} we apply the first-moment formalism to experimental spectra, imaging  entanglement in Olympecene graphene chains\cite{zhao25}.
Sec. \ref{sec:Mhigher} explores the additional information accessible through other moments, such as $n=2$ and $n=-1$.
Sec. \ref{sec:GTE} establishes connections between trimer excitation intensities and genuine tripartite entanglement, while
Sec. \ref{sec:quantum_circuit} presents quantum circuits that reproduce the spectroscopic information accessed by IETS, clarifying the non-local nature of spin-based IETS, in spite of the local nature of tunneling.
Sec. \ref{sec:disc} concludes with a discussion of the results and their experimental implications.

\begin{figure}[t]
\centering
\includegraphics[width=0.9\columnwidth]{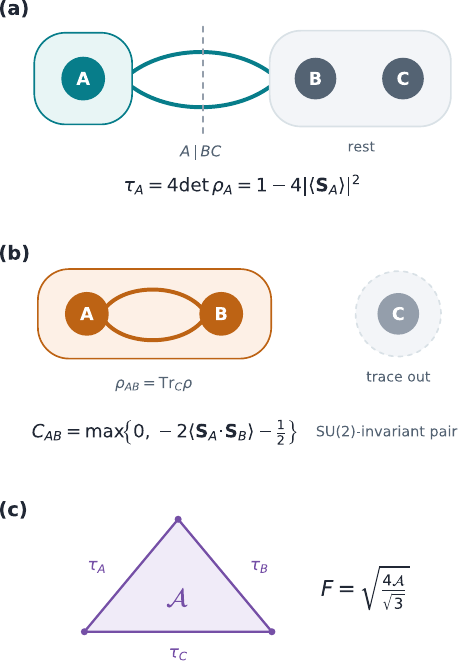}
\caption{
\textbf{Entanglement in spin-$1/2$ systems.} (a) The one-tangle $\tau_i$ quantifies entanglement between spin $i$ and the rest of a globally pure system. (b) The pairwise concurrence $C_{ij}$ is obtained from the two-spin reduced density matrix after tracing out all remaining spins; its square defines the two-tangle. (c) Genuine tripartite entanglement illustrated by the concurrence triangle, whose sides are the three one-tangles. 
}
\label{fig:entanglement_concepts}
\end{figure}

\section{Entanglement measures\label{sec:entanglement}}
In this section we review   the  three different entanglement measures that we have been able to  connect to the STM-based IETS. We distinguish entanglement between one spin and the remaining
system, pairwise entanglement, and genuine tripartite entanglement~\cite{amico08}. These three types of entanglement are schematically represented in Fig. \ref{fig:entanglement_concepts}).

\subsection{One spin versus the rest}
For a globally pure state $|\Psi\rangle$, the entanglement between a spin-$1/2$ degree of freedom $i$ and its complement $\bar{i}$ is characterized by the concurrence $C_{i|\bar{i}}$~\cite{rungta01}. Its square defines the one-tangle~\cite{coffman00},
\begin{equation}
\tau_i \equiv C_{i|\bar{i}}^2
=2\left[1-\operatorname{Tr}(\rho_i^2)\right]
=4\det\rho_i,
\end{equation}
where $\rho_i=\operatorname{Tr}_{\bar{i}} |\Psi\rangle\langle\Psi|$. Using the Bloch representation $\rho_i=[\boldsymbol{1} +2\langle\boldsymbol{S}(i)\rangle\cdot \boldsymbol{\sigma}]/2$, where $\boldsymbol{\sigma}$ are Pauli matrices, and $\langle \boldsymbol{S}(i) \rangle = \langle \Psi| \boldsymbol{S}(i) | \Psi \rangle$ is the expectation value of the local spin operator, gives
\begin{equation}
\tau_i=1-4|\langle\boldsymbol{S} (i) \rangle|^2.
\label{eq:one_tangle}
\end{equation}
If the system is separable across this bipartition, then $\tau_i=0$, whereas $\tau_i=1$ corresponds to maximal entanglement. These expressions apply to a pure ground state, but do not generally quantify entanglement for a thermal mixture. From Eq. (\ref{eq:one_tangle}) we can establish that the local magnetization and entanglement, through the one-tangle, are complementary quantities. 

As we show later in the text, STM-based IETS measured with the tip positioned over site $i$ provides access to $\tau_i$ in an interacting spin-$1/2$ system, enabling spatial imaging of the one-tangle.

\subsection{Pairwise entanglement}

The entanglement between two selected spins is determined by their reduced density matrix $\rho_{ij}=\operatorname{Tr}_{\overline{ij}}\rho$, which is generally mixed even when the full system is pure. Wootters's concurrence is~\cite{wootters98}
\begin{equation}
C_{ij}=\max\{0,\lambda_1-\lambda_2-\lambda_3-\lambda_4\},
\label{eq:wootters}
\end{equation}
where $\lambda_1\geq\lambda_2\geq\lambda_3\geq\lambda_4$ are the square roots of the eigenvalues of $\rho_{ij}\widetilde{\rho}_{ij}$, with
\begin{equation}
\widetilde{\rho}_{ij}
=(\sigma_y\otimes\sigma_y)\rho_{ij}^{*}
(\sigma_y\otimes\sigma_y).
\end{equation}
Here, complex conjugation is taken in the standard spin product basis. The concurrence is positive if and only if the two-qubit state is entangled; its square, $\tau_{ij}=C_{ij}^2$, is the two-tangle.

For a spin-rotational-invariant system, this expression simplifies to~\cite{oconnor01,amico08}
\begin{equation}
C_{ij}=
\max\left\{0,
-2\langle\boldsymbol{S}(i)\cdot\boldsymbol{S}(j)\rangle
-\frac12\right\}.
\label{eq:concurrence_correlator}
\end{equation}
where $\langle\boldsymbol{S}(i)\cdot\boldsymbol{S}(j)\rangle$ is the thermal expectation value of the spin correlator. Pairwise entanglement is therefore present precisely when $\langle\boldsymbol{S}(i)\cdot\boldsymbol{S}(j)\rangle<-1/4$. Hence, for an antiferromagnetically coupled spin-$1/2$ Heisenberg dimer with exchange coupling $J$ at zero magnetic field, thermal entanglement vanishes for $k_BT\geq J/\ln 3\simeq0.910\,J$.

\subsection{Genuine tripartite entanglement}
A pure three-qubit state is genuinely tripartite entangled if it is inseparable across all three bipartitions $A|BC$, $B|AC$, and $C|AB$. The corresponding one-tangles $\tau_A$, $\tau_B$, and $\tau_C$ define the side lengths of a concurrence triangle. Its normalized square-root area gives the concurrence fill~\cite{xie21},
\begin{equation}
F_{ABC}= \left[ \frac{16}{3}\, q(q-\tau_A)(q-\tau_B)(q-\tau_C) \right]^{1/4},
\label{eq:concurrence_fill}
\end{equation}
with 
\begin{equation}
q=\frac{\tau_A+\tau_B+\tau_C}{2}.
\end{equation}
For pure three-qubit states, $F_{ABC}$ vanishes for product and biseparable states and is positive for every genuinely tripartite entangled state. It thus provides a faithful indicator of genuine tripartite entanglement, although it is not an entanglement monotone under local operations and classical communication~\cite{ge23}. 

The result in Eq. (\ref{eq:concurrence_fill}) does not directly apply to a mixed three-spin reduced state embedded in a larger system, and a convex roof generalization is required in that case\cite{Bennett96}.  A particularly useful case where that generalization is trivial, is that of an isolated $S=1/2$ doublet state in a spin-rotational-invariant system, for example in a Heisenberg trimer. In the absence of a magnetic field, its zero-temperature equilibrium state is the unpolarized mixture
\begin{align}
    \rho_G=\frac{1}{2}\sum_{m=\pm1/2}
    |G,m\rangle\langle G,m|.
\end{align}
Every normalized pure state within this doublet is related to any other by a collective spin rotation. Since this rotation acts as a product of local unitaries on the three spins, the one-tangles are preserved (c.f. Eq. (\ref{eq:one_tangle})) and all such states have the same concurrence fill (see Eq. (\ref{eq:concurrence_fill})). Furthermore, every pure-state decomposition of $\rho_G$ contains only states within the doublet. Its convex-roof concurrence fill therefore satisfies
\[
    F_{\mathrm{cr}}(\rho_G)
    =F\bigl(|G,+1/2\rangle\bigr)
    =F\bigl(|G,-1/2\rangle\bigr).
\]
Consequently, the concurrence fill of the unpolarized ground-state mixture can be calculated using either representative pure ground state. At low but finite temperature, the equilibrium state is
well approximated by $\rho_G$ provided that $k_{\mathrm B}T\ll\Delta$, where $\Delta$ is the gap to the lowest excited multiplet.
 \

\subsection{Entanglement witnesses}
Entanglement witnesses establish the presence of entanglement without, in general, quantifying its amount. A witness operator $W$ satisfies $\operatorname{Tr}(\rho_{\mathrm{sep}}W)\geq0$ for every separable state; a negative expectation value therefore certifies entanglement. This is a sufficient, but not necessary, criterion, and applies to both pure and mixed states.

For interacting spins, the energy provides a natural witness~\cite{dowling04,toth05}. Defining the minimum energy compatible with a fully separable state,
\begin{equation}
    E_{\mathrm{sep}} =\min_{\rho_{\mathrm{sep}}} \operatorname{Tr}(\rho_{\mathrm{sep}}H),
\end{equation}
the condition $\langle H\rangle<E_{\mathrm{sep}}$ certifies entanglement. In particular, a ground-state energy $E_G<E_{\mathrm{sep}}$ establishes an entangled ground state; the interval between these energies is termed the entanglement gap~\cite{dowling04}.

Thermodynamic response functions provide complementary criteria. Specific heat, which probes energy fluctuations in thermal equilibrium, can likewise reveal entanglement through model-dependent bounds~\cite{wiesniak08}. Magnetic susceptibility can witness entanglement through bounds on collective spin fluctuations~\cite{wiesniak05}. Specifically, for $N$ spins of magnitude $S$ with spin-rotational invariance, the trace of the spin susceptibility  also provides an entanglement witness.

\section{Theory of inelastic electron tunnel spectroscopy\label{sec:IETS}}

\subsection{Perturbative theory of IETS}
Spin IETS probes magnetic excitations through the energy transferred by tunneling electrons to a surface-supported spin system. An excitation of energy $\Delta$ opens an additional transport channel at a bias $e|V|=\Delta$, producing a step in the 
conductance $G(V)=dI/dV$. The step intensity contains information about the spin matrix elements connecting the initial and final many-body states. This connection has been established using spin-assisted tunneling Hamiltonians and complementary scattering approaches~\cite{appelbaum67,fernandezrossier09,persson09,lorente09,ternes15}.

We consider unpolarized electrodes, approximately energy-independent electronic densities of states and tunneling amplitudes over the spectroscopic window, and a tip that predominantly couples to
spin $i$. To leading order in the tunneling amplitudes, the conductance can be decomposed as
\begin{equation}
    G_i(V)=G_{{\rm ns},i}+G_{{\rm s,el},i}+G_{{\rm s,inel},i}.
\label{eq:Gdecomposition}
\end{equation}
Here, $G_{{\rm ns},i}$ is the spin-independent background, whereas the remaining terms describe elastic, $G_{{\rm s,el},i}$, and inelastic, $G_{{\rm s,in},i}$, spin-assisted tunneling. 

The  expressions for $G_{{\rm s,el},i}+G_{{\rm s,inel},i}$ used below are obtained as  the leading nonvanishing term in a perturbative expansion in the  Kondo couplings between the localized spins and the tip and substrate electrons~\cite{appelbaum67,fernandezrossier09,ternes15}. At this order, the spin eigenstates  are calculated for the isolated interacting spin system: entanglement among the surface spins is retained nonperturbatively, whereas their entanglement with the electrode electrons is neglected.  The resulting description applies when electrode-induced
screening, level renormalization and lifetime broadening remain small on the energy scales relevant to the measurement.

For equilibrium occupations $p_m=e^{-\beta E_m}/Z$, with $\beta=k_BT$ and $Z$ the partition function, the spin-assisted conductance, as a function of the applied bias  is~\cite{fernandezrossier09,ternes15} 
\begin{equation}
G_{{\rm s},i}(V)=g_i\sum_{m,n}p_m W_{mn}^{(i)}\,{\cal K}_T(V,E_n-E_m),
\label{eq:GIETS}
\end{equation}
where $g_i$ is a conductance prefactor containing the tunneling amplitude and electrode densities of states. The thermal kernel is
\begin{equation}
\begin{split}
{\cal K}_T(V,\Delta)
&=\Theta\!\left(\frac{eV-\Delta}{k_BT}\right)
 +\Theta\!\left(\frac{-eV-\Delta}{k_BT}\right),\\
\Theta(x)&=\frac{d}{dx}\left(\frac{x}{1-e^{-x}}\right),
\end{split}
\label{eq:IETSkernel}
\end{equation}
with $\Theta(0)=1/2$ defined by continuity. Equation~\eqref{eq:GIETS} includes both elastic and inelastic contributions. For an isotropic exchange interaction between the tunneling electron and spin $i$, the transition weight is\cite{fernandezrossier09}
\begin{equation}
W_{mn}^{(i)}
=\sum_{a=x,y,z}
\left|\langle n|S_a(i)|m\rangle\right|^2,
\label{eq:spinweight}
\end{equation}
where $|m\rangle$ and $|n\rangle$ are eigenstates of the interacting spin Hamiltonian, with energies $E_m$ and $E_n$. We use dimensionless spin operators throughout. The sum over spin components follows from the isotropic tunneling exchange and the trace over the unpolarized electron spins; it does not require the spin Hamiltonian itself to be rotationally invariant.

For a nondegenerate ground state $|G\rangle$ at zero temperature, each excited state $|n\rangle$ contributes a conductance step at $e|V|=E_n-E_G$, with height $g_iW_{Gn}^{(i)}$. The elastic
contribution is instead proportional to $\sum_a|\langle G|S_a(i)|G\rangle|^2$.

This description provides the starting point for the spectroscopic sum rules developed below. It accounts for the conductance steps~\cite{hirjibehedin06,chen08, spinelli14, Mishra21, zhao25}
that appear when the bias reaches the threshold for a transition between spin eigenstates connected by a nonzero spin matrix element, $|eV|=E_n-E_m$. The step heights encode both the occupation
$p_m$ of the initial state, typically the ground state at low temperature, and the corresponding spin transition matrix elements. In extended spin structures, these matrix elements depend on the
tip position, giving rise to spatial variations in the step heights that enable imaging of the spin excitations~\cite{spinelli14,Mishra21,zhao25}. The theory assumes thermal occupations $p_m$, which are justified when spin relaxation is fast compared with the mean time between successive tunneling events, so that the spin system returns to thermal equilibrium between events~\cite{delgado10}. This regime can be reached by retracting the tip to reduce the tunneling current and thereby suppress current-induced changes in the state occupations.

\subsection{A simple connection between IETS and entanglement}
We start to build our case that IETS can provide information about entanglement by considering the simplest case of a $S=1/2$ dimer, with arbitrary Hamiltonian.  We show now that 
if the number of inelastic conductance steps is larger than two, at least one of the four eigenstates of the Hamiltonian must be entangled.  Hence, we establish the number of inelastic steps in the IETS as an entanglement witness. In a similar fashion to other entanglement witnesses, a number of steps smaller or equal to this threshold value does not allow us to infer anything about entanglement in the system.

We shall prove the above statement by contradiction. Assume that the four eigenstates of the arbitrary Hamiltonian form an orthonormal product basis,
\begin{equation}
|n\rangle=|a_n\rangle_A\otimes|b_n\rangle_B, \quad n=0,1,2,3,
\end{equation}
and that all three transitions from the only occupied state, $|0\rangle$, to states $|1\rangle$, $|2\rangle$, and $|3\rangle$ are allowed when the tip probes spin $A$. For each transition, at least one spin component $\alpha=x,y,z$ must have a nonzero matrix element,
\begin{equation}
\langle n|S_\alpha(A)|0\rangle = \langle a_n|S_\alpha(A)|a_0\rangle \langle b_n|b_0\rangle\neq0.
\end{equation}
Hence, $\langle b_n|b_0\rangle\neq0$ for $n=1,2,3$. Orthogonality to the initial state then gives
\begin{equation}
0=\langle n|0\rangle =\langle a_n|a_0\rangle\langle b_n|b_0\rangle \quad\Longrightarrow\quad \langle a_n|a_0\rangle=0.
\end{equation}
Since spin $A$ has a two-dimensional Hilbert space, all three excited states must therefore share the same local factor $|a_0^\perp\rangle$, up to phases that can be absorbed into the states of spin $B$:
\begin{equation}
|n\rangle=|a_0^\perp\rangle_A\otimes|b_n\rangle_B, \qquad n=1,2,3.
\end{equation}
Their mutual orthogonality consequently requires
\begin{equation}
\langle b_1|b_2\rangle =\langle b_1|b_3\rangle =\langle b_2|b_3\rangle=0.
\end{equation}
This is impossible: the two-dimensional Hilbert space of spin $B$ cannot accommodate three mutually orthogonal states. Thus, at least one transition must be forbidden if all eigenstates are product states. The same argument applies upon exchanging $A$ and $B$.

Observation of three distinct inelastic thresholds therefore rules out a complete product eigenbasis and certifies the presence of entangled eigenstates, although it does not by itself establish that the initial state is entangled. This witness assumes a single occupied initial state and a tunnelling operator acting on only one spin.
We can thus be sure that entanglement is present in at least one of the four states of the   antiferromagnetically coupled  $S=1/2$ cobalt phthalocyanine molecules in stacks on Pb, where a three-threshold spectrum was reported in Ref. ~\cite{chen08}.

\section{Zeroth moment and the one-tangle \label{sec:M0}}

In the following sections we shall present a general moment decomposition of the differential conductance:
\begin{align}
    \mathcal{M}^{(n)}(i) = \int_0^\infty V^n \frac{d^2I(i)}{dV^2}dV \label{eq: moment def}
\end{align}
and establish its relation to the one-tangle, spin correlators, concurrence, and energy-based witnesses. In this section we focus on the specific case of $n=0$.

\subsection{Theory}

The zeroth moment of the local IETS spectrum measures the total inelastic spin spectral weight. Within the conductance decomposition 
introduced above in Eqs. (\ref{eq:Gdecomposition})-(\ref{eq:GIETS}), the zeroth moment is given by
\begin{equation}
    \mathcal{M}^{(0)}(i)\equiv \int_0^\infty dV\,\frac{dG_{s,i}}{dV} =G_{s,i}(\infty)-G_{s,i}(0),
\label{eq:zeroth_moment}
\end{equation}
where $G_{s,i}(\infty)$ denotes the conductance plateau above all spin excitation thresholds, within the validity range of the spin-only description.

For biases several thermal energies above the highest excitation threshold, the thermal kernel of Eq. (\ref{eq:IETSkernel}) approaches unity. Hence, Eq. (\ref{eq:GIETS}) can be simplified to\cite{loth10}:
\begin{align}
    G_{s,i}(\infty) &=g_i\sum_{m,n} p_m\sum_{a=x,y,z} |\langle n|S_a(i)|m\rangle|^2 \notag \\
                    &=g_i\sum_m p_m \langle m|\boldsymbol{S}^2(i)|m\rangle=g_iS_i(S_i+1).
\end{align}
Thus, the high-bias conductance, including the spin-independent background, is $G_i(\infty)=G_{\mathrm{ns},i}+g_iS_i(S_i+1)$, for arbitrary local spin $S_i$.

We next consider a unique ground state $|G\rangle$, separated from the excited states by a gap $\Delta$, with negligible thermal depletion, $k_BT\ll\Delta$.
At zero bias, inelastic transitions are suppressed and only the elastic channel remains. Consequently,
\begin{equation}
    G_{s,i}(0) =g_i\sum_{a=x,y,z} |\langle G|S_a(i)|G\rangle|^2 =g_i|\langle\boldsymbol{S}(i)\rangle|^2. \label{eq: Gsi_0=magnetization}
\end{equation}
Subtracting the zero-bias conductance from the high-bias plateau therefore yields:
\begin{equation}
    \frac{\mathcal{M}^{(0)}(i)}{g_i} =S_i(S_i+1) -|\langle \boldsymbol{S}(i) \rangle|^2.
\label{eq:zeroth_moment_unique}
\end{equation}
There are two important aspects which are worth emphasizing about this expression. First, the zeroth moment relates to the sum of the variances of the three components of the local spin operator:
\begin{align}
    \sum_{a} \operatorname{Var}_G[S_a(i)] &=\sum_{a} \left[ \langle G|S_a^2(i)|G\rangle -\langle G|S_a(i)|G\rangle^2 \right]\nonumber \\
    &= S_i(S_i+1)-|\langle \boldsymbol{S}(i)\rangle|^2
\end{align}
Second, Eq. (\ref{eq:zeroth_moment_unique}) connects the height of the inelastic steps to the length of the average magnetization vector, without making use of a spin-polarized tip. 
If instead of a unique ground state, we consider a degenerate manifold $\mathcal{G}$ (which still exhausts the thermal occupations), Eq. (\ref{eq:zeroth_moment_unique}) generalizes to:
\begin{equation}
    \frac{\mathcal{M}^{(0)}(i)}{g_i} =S_i(S_i+1) -\sum_{g,g'\in\mathcal{G}}p_g \sum_{a=x,y,z} |\langle g'|S_a(i)|g\rangle|^2.
    \label{eq:zeroth_moment_degenerate}
\end{equation}

For interacting spin-$1/2$ systems with a unique ground state, the relation between local magnetization and the one-tangle introduced in Sec.~\ref{sec:entanglement} then yields
\begin{equation}
    \frac{\mathcal{M}^{(0)}(i)}{g_i} =\frac{3}{4} -|\langle\boldsymbol{S}(i)\rangle|^2 =\frac{2+\tau_i}{4},
\label{eq:main1}
\end{equation}
or, equivalently,
\begin{equation}
\tau_i=4\frac{\mathcal{M}^{(0)}(i)}{g_i}-2.
\label{eq:one_tangle_iets}
\end{equation}
The normalized zeroth moment therefore directly quantifies entanglement between spin $i$ and the remaining system when the global state is pure. This identification does not generally extend to a statistical mixture within a degenerate ground-state manifold.  Eq. (\ref{eq:main1}) is one of the main results of this work. 

In the following subsections we discuss the results we have just presented could be used in existing and future experiments.

\begin{figure}[t]
    \includegraphics[width=0.5 \textwidth]{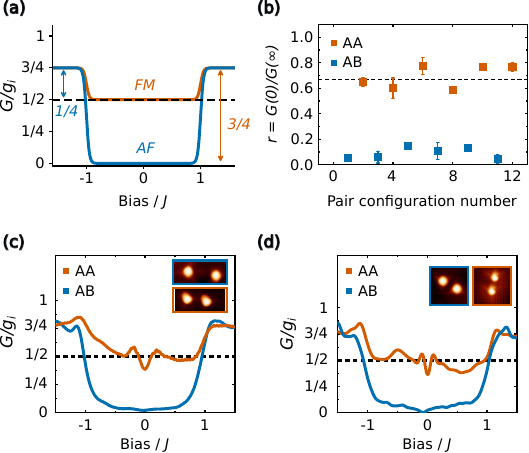}
    \caption{
    \textbf{Inelastic step heights in antiferromagnetic and
    ferromagnetic spin-$1/2$ dimers.}
    (a) Ideal conductance spectra for antiferromagnetic (AFM, blue) and ferromagnetic (FM, orange) Heisenberg dimers. In units of the spin-dependent conductance prefactor, the inelastic step heights are $3/4$ and $1/4$, respectively, while the high-bias conductance is $3/4$ in both cases.
    (b) Experimental ratio $r=G(0)/G(\infty)$ for hydrogen pairs occupying opposite graphene sublattices (AB) or the same sublattice (AA). The dashed line indicates the ideal ferromagnetic value $r=2/3$.
    (c,d) Representative experimental spectra for AB and AA pairs, with bias scaled by the magnitude of the exchange coupling. Insets show the corresponding STM images.
    Insets show STM topography images of each pair configuration (AB:3.9x9.0 nm2, 100 mV, 50 pA, AA: 3.4x7.2 nm2, 50 mV, 300 pA) and (AB: 6.4x6.4 nm2, 200 mV, 100 pA, AA:6.3x6.3 nm2, 10 mV, 50 pA), for panels (c) and (d), respectively.
    }
    \label{fig:AA_AB_step_height}
\end{figure}

\subsection{Ferro- vs Antiferromagnetic Heisenberg spin-1/2 dimers}
Let us first consider the case of Heisenberg spin-1/2 dimers, with either ferro- or antiferromagnetic exchange coupling. For otherwise identical tunneling conditions, our theory predicts that the exchange-induced inelastic step of an antiferromagnetic spin-$1/2$ Heisenberg dimer is three times larger than that of its ferromagnetic counterpart: the singlet-to-triplet transitions carry total weight $3/4$, whereas the triplet-to-singlet transitions carry weight $1/4$ (see Fig. \ref{fig:AA_AB_step_height}a). This is expected from the previous subsection, since in both cases the spin conductance saturates at $G_{s,i}(\infty) = 3/4$, while the spin-elastic contribution is $G_{s,i}(0)=0$ for the antiferromagnetic dimer (since it is proportional to the local magnetization, which vanishes in a singlet), and $G_{s,i}(0)=1/2$ for the ferromagnetic one, following Eq. (\ref{eq:zeroth_moment_degenerate}). 

In Fig.~\ref{fig:AA_AB_step_height} we compare these predictions with experimental spectra of dimers of atomic hydrogen chemisorbed on graphene~\cite{vina26}. Panel~(b) summarizes the zero-bias to high-bias conductance ratio for 12 dimers, while panels~(c) and (d) show four representative spectra, two for AB pairs and two for AA pairs. Near charge neutrality, the vanishing density of states of graphene suppresses the spin-independent elastic contribution, making the spin-dependent conductance more prominent. 

Antiferromagnetically coupled opposite-sublattice (AB) pairs exhibit nearly vanishing zero-bias conductance, whereas ferromagnetically coupled same-sublattice (AA) pairs retain a finite conductance. We attribute the latter to elastic spin scattering within the degenerate ground-state manifold, described by the second term of Eq.~\eqref{eq:zeroth_moment_degenerate}. The smaller inelastic conductance contrast of AA pairs is qualitatively consistent with our prediction, although  the experimental data shows signals close to zero bias which are not accounted for by the simulation.

%

\subsection{Locally biased Heisenberg $S=1/2$ dimers}

We now propose an experiment, feasible with state of the art, where the entanglement of an antiferromagnetically spin-1/2 pair can be changed, in a control manner, and predict how the IETS lineshape would change.

\begin{figure}[t]
\centering
\includegraphics[width=0.45\textwidth]{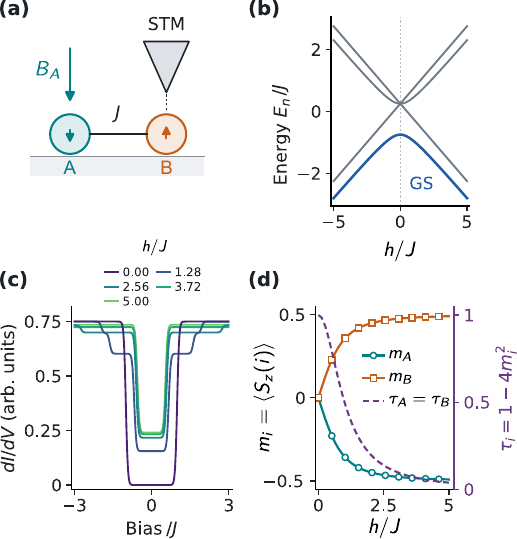}
\caption{
\textbf{Entanglement and local IETS of an antiferromagnetic spin-$1/2$
dimer.}
(a) Schematic: the STM probes spin $B$, while a local field
acts on spin $A$.
(b) Eigenenergies as a function of the local Zeeman energy $h$.
(c) Spin-dependent differential conductance at site $A$
for different fields, calculated at $T=300$~mK with $J=1$meV.
The spin-independent background is omitted.
(d) Ground-state one-tangle $\tau_A=\tau_B$ and local
magnetizations $\langle S_z(A)\rangle$ and $\langle S_z(B)\rangle$.
}
\label{fig:dimer_local_field}
\end{figure}

We consider a $S=1/2$ Heisenberg dimer, with antiferromagnetic exchange $J$, and a magnetic field that acts only on one of the spins. This local field could be originated by exchange coupling to an Ising spin nearby\cite{wang23}. Assuming the field couples to site $A$ only, we write:
\begin{equation}
    H=J\,\boldsymbol{S}(A)\cdot\boldsymbol{S}(B)-hS_z(A), \; J>0,\; h=g\mu_B B_A,
\label{eq:dimer_local_field}
\end{equation}
where the spin operators are dimensionless. A schematic of the system is given in Fig. \ref{fig:dimer_local_field}a.

At $h=0$, the ground state is the maximally entangled singlet, with vanishing local magnetization, maximal zeroth moment, and one-tangle equal to 1. As we ramp-up $h$ the ground state remains in the $S_z^{\mathrm{tot}}=0$ sector, and its wave function reads:
\begin{equation}
|G(h)\rangle=\frac{1}{\sqrt{2}}\left(
\sqrt{1+x}\,|\downarrow\uparrow\rangle
-\sqrt{1-x}\,|\uparrow\downarrow\rangle\right),
\label{eq:ansatz}
\end{equation}
where $x=h/\sqrt{J^2+h^2}$. Therefore, we see that as $h$ increases, the two amplitudes become unequal and the state approaches $|\downarrow\uparrow\rangle$ in the limit $h/J\rightarrow\infty$. As a consequence, local magnetizations increase in magnitude accompanied by a growth in the zero bias conductance (c.f. Eq. (\ref{eq: Gsi_0=magnetization})); at the same time, entanglement, together with the zeroth moment, decrease (c.f. Eq. (\ref{eq:main1})).
Figure~\ref{fig:dimer_local_field} illustrates this connection through the evolution of the excitation spectrum, the conductance steps, and the ground-state one-tangle and magnetizations. We note that, for the values of $J=1$meV and $T=300$mK used in the simulation the ground-state occupation exceeds $1-10^{-9}$ throughout the field range shown. Thermal corrections are therefore negligible, and the observed evolution reflects ground-state properties.

The same analysis can be extended to the emergent edge spins of a topological spin chain  \cite{delCastillo26}, such as those made with nanographenes\cite{zhao24} or Ti atoms on MgO\cite{wang24}. Application of a local field in one side of the chain also drives the ground state from an entangled state, with vanishing local magnetization and maximal one-tangle in the entire chain, to a product state where the edges become magnetized and the one-tangle is consequently reduced there.  Therefore, STM-based IETS could be used to image entanglement in this class of system through the on-tangle, via Eq. (\ref{eq:main1}).

\section{First moment and the concurrence \label{sec:M1}}

\subsection{Theory}
In the previous section we have demonstrated that the 0th moment provides valuable information about a spin system, namely through  the relation between the one-tangle and the local magnetization. Now, we move on to the next spectral moment, $n=1$, to explore what relevant quantities can be produced in this case.

Integrating by parts, Eq. (\ref{eq: moment def}) can be written for $n=1$ as:
\begin{align}
    \mathcal{M}^{(1)}(i) = \int_0^\infty G_{\textrm{s},i}(\infty) - G_{\textrm{s},i}(V)dV,
\end{align}
which is simply the area between the saturation plateau and the $dI/dV$ curve and its saturation value. Explicitly performing the integral, one finds: 
\begin{align}
    \frac{\mathcal{M}^{(1)}(i)}{g_i} &=\frac{1}{2}\sum_m p_m \sum_{a=x,y,z} \langle m| S_a(i) (H-E_m) S_a (i) |m\rangle \notag \\
    &= \frac{1}{2}\sum_m p_m \sum_{a=x,y,z} \langle m| [S_a(i), [H, S_a(i)]]|m\rangle \label{eq: 1st moment}
\end{align}
which is valid for any temperature $T$, and where $H$ is the spin Hamiltonian with $H|m\rangle = E_m |m\rangle$. One direct physical interpretation of this result, is that we can identify the first moment at a given site $i$ as a local exchange-energy stiffness, i.e. measurement of the energy cost of infinitesimally rotating a local spin. To see this, we start by defining $E(a,i,\theta) = \langle U_a^\dagger(i,\theta) H U_a(i,\theta)\rangle$, where $U_a(i,\theta)=\textrm{exp}(-i\theta S_a(i))$ is the unitary rotation operator of spin $i$. Differentiating $E(a,i,\theta)$ twice relative to $\theta$ yields:
\begin{align}
    \frac{\mathcal{M}^{(1)}(i)}{g_i} =\frac{1}{2} \sum_{a=x,y,z} \frac{d^2E(a,i,\theta)}{d\theta^2}\bigg|_{\theta=0},
\end{align}
therefore establishing the relation between the first moment and a local exchange-energy stiffness.

We now consider the particular case where $H$ is a Heisenberg Hamiltonian with arbitrary exchange couplings, such that $H = \sum_{ij} J_{ij} \boldsymbol{S}(i) \cdot \boldsymbol{S}(j)$. Inserting this in Eq. (\ref{eq: 1st moment}), one finds:
\begin{align}
    \frac{\mathcal{M}^{(1)}(i)}{g_i} = - \sum_{j \in \mathcal{N} (i)} J_{ij} \langle \boldsymbol{S}(i) \cdot \boldsymbol{S}(j) \rangle \label{eq: 1st moment bilinear}
\end{align}
with $\mathcal{N}(i)$ labeling all the sites which are exchange coupled to site $i$, and $\langle ...\rangle$ encoding the thermal expectation value of the product of spin operators. In fact, a very similar result is found for any bilinear Hamiltonian, $H = \sum_{ij} J^{ab}_{ij} S_a(i) S_b(j)$ (with repeated indices summed over), yielding $\mathcal{M}^{(1)}(i) = - \sum_{j \in \mathcal{N} (i)} J_{ij}^{ab} \langle S_a (i) S_b (j) \rangle$, instead. Relations between magnetic spectral moments and exchange correlations have a long history in neutron-scattering theory~\cite{hohenberg74,zheludev04}.
For simplicity's sake, we focus the rest of the analysis on the simpler Heisenberg model, with the extension to the more general case being straightforward. 

The result in Eq. (\ref{eq: 1st moment bilinear}) is one of the central ones in our work, and it establishes a relation between the first moment and the thermal spin correlators that, for spin rotationally invariant systems, relate to the Wootters concurrence of Eq. (\ref{eq:concurrence_correlator}). Summing $\mathcal{M}^{(1)}(i)$ over all sites leads to an expression relating the first moments to the internal energy of the spin Hamiltonian, $U(T)$:
\begin{align}
    \sum_i \frac{\mathcal{M}^{(1)}(i)}{g_i} = -2U(T). \label{eq: 1st moment EGS}
\end{align}
where the factor of $2$ comes from double counting over sites, and the dependence on temperature was made explicit.

Evaluating the first moments from $d^2I/dV^2$ obtained at different temperatures, it is possible to obtain resolved spectroscopic access to the thermal redistribution of exchange energy, together with the total magnetic heat capacity:
\begin{align}
    C_V(T)= \frac{dU}{dT} = -\frac{1}{2} \sum_i \frac{d \mathcal{M}^{(1)}(i,T)}{dT}, \label{eq: 1st moment Cv}
\end{align}

We now emphasize that including other types of interactions in the Hamiltonian, such as single-spin terms (e.g., a Zeeman term) or biquadratic interactions, prevents the reconstruction of the internal energy by simply summing all first moments. The nested commutator in Eq. (27) produces terms of the same form as those appearing in the Hamiltonian: bilinear interactions lead to bilinear contributions, biquadratic interactions to biquadratic contributions, and so on. However, each type of interaction enters with a different numerical prefactor, depending on the number and structure of the spin operators involved. Consequently, summing the first moments no longer directly yields the system's internal energy.

The result in Eq. (\ref{eq: 1st moment bilinear}) is valid for any arbitrary exchange connectivity of the spin Hamiltonian. However, for particular cases of the underlying coupling graph, additional information can be extracted from the first moments. If the connectivity of the Hamiltonian is such that the sites which are exchange coupled form a tree graph, where there exists a unique path between any two vertexes, e.g. a chain with first neighbor exchange couplings only, then the first moments determine every bond-energy contribution $B_{ij} = J_{ij} \langle \boldsymbol{S}(i)\cdot \boldsymbol{S}(j) \rangle$. This is possible because a tree graph has no cycles, and therefore all the $B_{ij}$ can be obtained recursively starting from terminal spins. 

Being even more restrictive, if the exchange couplings $J_{ij}$ are known, and we consider a system with nearest neighbor coupling only, the spin-spin correlators $\langle \boldsymbol{S}(i)\cdot \boldsymbol{S}(i+1) \rangle$ can be determined, since in that case the system of equations relating the first moments and the correlators can be inverted. Therefore, in such a case, the first moments would allow to image Wooters entanglement in all the bonds of the chain. This result is exemplified in the next subsection. Unlike the relation between the zeroth moment and the one-tangle, this result is valid even if the ground state is not a global pure state, although spin-rotational invariance is required.

Even when the exchange couplings are not known \emph{a priori}, we can still determine how the correlators evolve with temperature, under the assumption that the exchange couplings themselves are temperature independent:
\begin{align}
    \frac{B_{ij}(T_1)}{B_{ij}(T_2)} = \frac{\langle \boldsymbol{S}(i) \cdot \boldsymbol{S} (j)\rangle _{T_1}}{\langle \boldsymbol{S}(i) \cdot \boldsymbol{S} (j)\rangle _{T_2}}.
\end{align}
This is particularly interesting near low-energy magnetic rearrangements, thermally activated edge states, or the onset of triplet population in weakly coupled dimers. 

Finally, we note that we can use the first moment to test assumptions about the spin Hamiltonian. Suppose the exchange graph is bipartite, with sublattices $A$ and $B$. Then:
\begin{align}
    \sum_{i\in A}\mathcal{M}^{(1)}(i) = -\sum_{i\in A} \sum_{j\in \mathcal{N}(i)} B_{ij},
\end{align}
and likewise for the other sublattice
\begin{align}
    \sum_{j\in B}\mathcal{M}^{(1)}(j) = -\sum_{j\in B} \sum_{i\in \mathcal{N}(j)} B_{ij},
\end{align}
which leads to the following relation:
\begin{align}
    \sum_{i\in A}\mathcal{M}^{(1)}(i) - \sum_{j\in B}\mathcal{M}^{(1)}(j) = 0.
\end{align}
A violation can indicate interactions beyond the proposed bipartite exchange graph. This approach can be straightforwardly applied to validate  Hamiltonian learning in atomically engineered quantum magnets~\cite{lupi25}. In particular, the first-moment sum rules allow us to test the consistency of an inferred Hamiltonian and identify interactions missing from the assumed model.

\subsection{Applications}

\begin{figure}
    \centering
    \includegraphics[width=\linewidth]{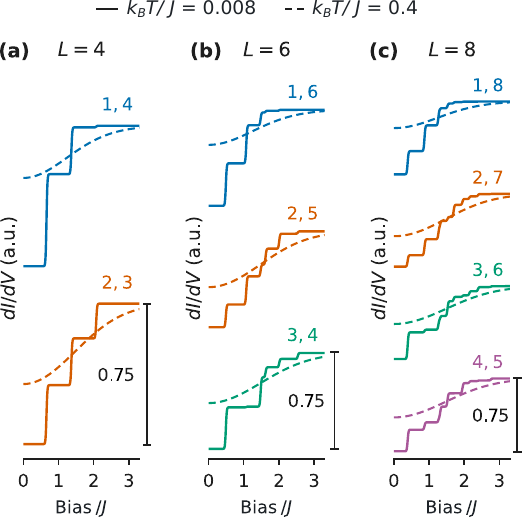}
    \caption{\textbf{Site-resolved differential conductance of open spin-$1/2$ Heisenberg chains.} Panels (a), (b) and (c) show simulations for chains with length $L=4,6$ and $8$ respectively, at low (solid) and high (dashed) temperature, with $k_BT/J=0.008$ and $k_BT/J=0.4$, respectively. Labels identify reflection-equivalent sites. Positive-bias spectra are vertically offset.}
    \label{fig:dI/dVs AF chain}
\end{figure}

To put to the test the results we have just derived, we are going to apply them to the case of a homogeneous nearest neighbor spin-1/2 Heisenberg chain:
\begin{align}
    H = J \sum_{i=0}^{L-1} \boldsymbol{S}(i) \cdot \boldsymbol{S}(i+1) \label{eq: AF Heisen H}.
\end{align}
Pairwise entanglement in finite Heisenberg chains has been studied explicitly, including its temperature and magnetic-field dependence~\cite{arnesen01}, as well as and the spatial modulation of ground-state concurrence induced by open boundaries~\cite{wang04}. Here, we consider shorter chains, with lengths $L=4,6,8$, relevant for the discussion of section \ref{sec:experimental} where we extract the first-neighbour concurrences out of experimental data. Below, we perform simulations at different temperatures and we compute conductances using the expressions given in Sec. \ref{sec:IETS}. The results are displayed in Fig. \ref{fig:dI/dVs AF chain}. There we see that for each chain, inelastic excitation steps appear at different bias, corresponding to the excitations of the spin model. The height of these steps is modulated from site to site, and increasing temperature increases their smearing.

\begin{figure}
    \centering
    \includegraphics[width=\linewidth]{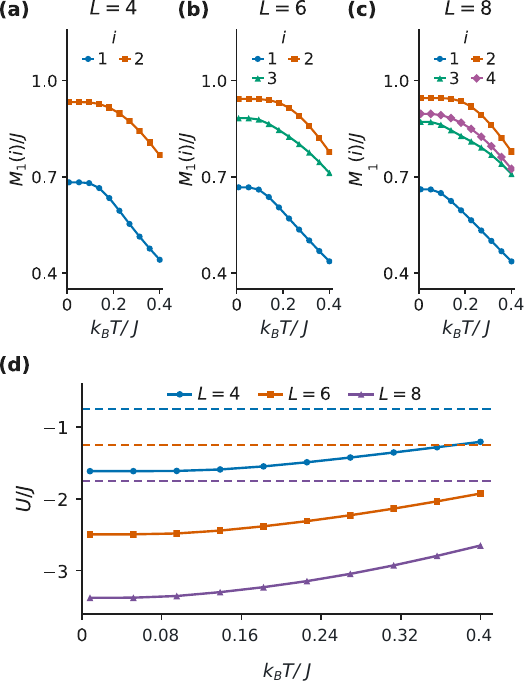}
    \caption{\textbf{First moments and internal energy as a function of temperature.} (a–c) Temperature dependence of local first moments for $L=4,6,8$, respectively. Legends give representative site indices $i$, which are equivalent to sites $L+1-i$. (d) Total internal energies $U$ (solid) and fully separable bounds $U_{\mathrm{sep}}=-(L-1)J/4$ (dashed, matching colors). Energies below their corresponding bounds certify entanglement.}
    \label{fig:internal energy AF chains}
\end{figure}

From the simulated $dI/dV$, we compute the first moments, and from there extract several meaningful quantities. The $\mathcal{M}^{(1)}(i)$ for the three chain lengths are depicted as a function of temperature in the top panels of Fig. \ref{fig:internal energy AF chains}. The first figure of merit we look at is the internal energy, $U$, and how it can be used as an entanglement witness. This is achieved by first computing the classically separable value for $U$, and then comparing it with the value obtained through the sum of all first moments. If the former is bigger than the latter, then we ascribe the energy difference to entanglement. The classical estimate for $U$ is obtained by treating the spins as rigid vectors, oriented in a Néel configuration, in which case one find $U_{\textrm{sep}} = - J(L-1)/4$. In Fig. \ref{fig:internal energy AF chains}d we show the evolution of the ground state energy with temperature, and find that for the three chains, the ground state energy increases with temperature, but it always stays below the classical separable limit, certifying entanglement, even at high temperatures (with $k_B T > 0.4J$). The result obtained from the $dI/dV$ through the first moment, matches exactly whith the result found from exact diagonalization of the spin model.

\begin{figure}
    \centering
    \includegraphics[width=\linewidth]{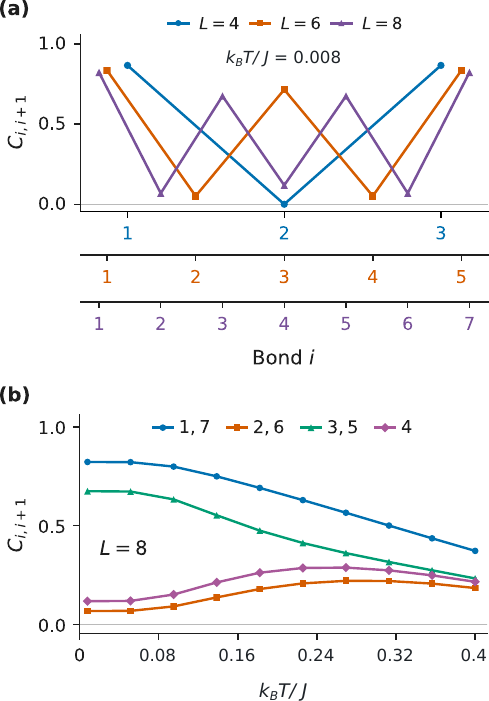}
    \caption{\textbf{Site resolved concurrence and its temperature dependence. }(a–c) Nearest-neighbor concurrence at low temperature ($k_BT/J=0.008$) for $L=4,6,8$, respectively. Bond $i$ denotes the pair $(i,i+1)$. (d) Temperature dependence for $L=8$; labels identify reflection-equivalent bonds.}
    \label{fig:bond energies AF chains}
\end{figure}

Since the nearest neighbor chain is a tree graph, all bond energies can be recovered recursively starting from the edge spins. Furthermore, since we also know the exchange coupling used in the simulations, the spin-spin correlators can be retrieved. Since the system we are considering respects spin-rotational invariance, the concurrence follows directly from Eq. (\ref{eq:concurrence_correlator}). These results are summarized in Fig. \ref{fig:bond energies AF chains}, where we find that, despite the homogeneous exchange, the open boundary conditions impose a strong modulation of the concurrence $C_{i,i+1}$ (and therefore correlators) across the chains. When studying the dependence of the concurrence with temperature (see bottom panel of Fig. \ref{fig:bond energies AF chains}), we find that at large temperatures, all correlators decrease, and the system becomes progressively less entangled. However, there is an intermediate temperature region, where a temperature increase actually produces an entanglement increase on the even-numbered bonds. As before, these result perfectly match those obtained from the direct diagonalization of the spin model.

\section{Application to experimental data} \label{sec:experimental}
We now apply the first-moment formalism to experimental tunnelling spectra to reconstruct the spatial distribution of pairwise entanglement. We consider the nanographene chains studied in Ref.~\cite{zhao25}, which realize the antiferromagnetic spin-$1/2$ Heisenberg Hamiltonian of Eq.~(\ref{eq: AF Heisen H}). Specifically, we analyze chains of lengths $L=4,6,8$, using their spatially resolved $dI/dV$ spectra and the independently determined nearest-neighbour exchange coupling $J=38\,\mathrm{meV}$.

In Secs.~\ref{sec:M0} and \ref{sec:M1}, the moments were defined after dividing out the tunnelling prefactor $g_i$. Although this prefactor can be set to unity in simulations, its experimental value must be determined through calibration. One possible reference is an antiferromagnetic spin-$1/2$ dimer. At temperatures well below its singlet--triplet gap, the normalized zeroth moment at either site is $s(s+1)=3/4$. The measured zeroth moment therefore determines the corresponding tunnelling prefactor, $g_i$. This calibration can be transferred to the target system provided that the effective prefactor remains unchanged between the reference and target measurements.

Alternatively, calibration can be performed directly on the target system when its ground state is a nondegenerate singlet and thermal excitations are negligible. In this regime, the spin-dependent conductance vanishes at zero bias, and the normalized zeroth moment is again $3/4$. ~The appropriate temperature criterion is $k_{\mathrm B}T\ll\Delta_L$, where $\Delta_L$ is the lowest excitation gap of the chain, rather than simply $k_{\mathrm B}T\ll J$. This distinction becomes increasingly relevant in systems where the gap decreases with chain length.

For the $L=4,6,8$ spectra considered here, the spin-dependent conductance is negligible at zero bias, consistent with the low-temperature singlet regime. Knowing the bias window captures all the relevant spin-excitation weight \cite{zhao25}, we determine $g_i$ separately at each measurement position by imposing a normalized zeroth moment of $3/4$. We then evaluate the normalized first moments and reconstruct the nearest-neighbour spin correlators using the first-moment relations for the open Heisenberg chain and the known value of $J$.

\begin{figure*}
    \centering
    \includegraphics[width=\linewidth]{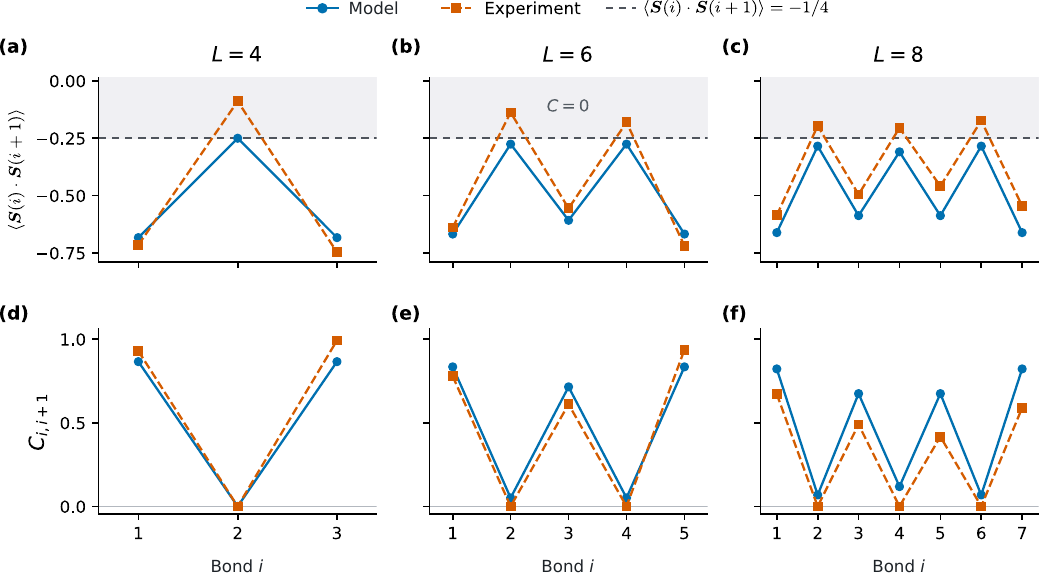}
 \caption{\textbf{Nearest nerighbor correlators and concurrence extracted from experimental data.} 
Nearest-neighbour spin correlators (a--c) and pairwise
concurrence (d--f) for open spin-$1/2$ Heisenberg chains
with $L=4,6,8$, respectively.
Blue circles show the ground-state predictions, while
orange squares show the values reconstructed from the
experimental IETS spectra for olimpecene chains\cite{zhao25}.
The dashed horizontal line at
$\langle\boldsymbol{S}(i)\cdot\boldsymbol{S}(i+1)\rangle=-1/4$
marks the threshold for pairwise entanglement:
correlators below this line give finite concurrence,
whereas the shaded region corresponds to $C_{i,i+1}=0$.
The concurrences  are obtained from
the correlators using Eq.~\eqref{eq:concurrence_correlator}.
Bond $i$ connects sites $i$ and $i+1$.
Lines are guides to the eye.
}   
    \label{fig:exp_vs_theory_Chenxiao}
\end{figure*}

As discussed before, for an SU(2)-invariant state of spin-$1/2$ particles, the spin-spin correlators determine the nearest-neighbour concurrence through Wooter's formula (c.f. Eq. (\ref{eq:concurrence_correlator})). Figure~~\ref{fig:exp_vs_theory_Chenxiao} compares the spin correlators and concurrence reconstructed from experiment with the ground-state predictions of the same Heisenberg model. The reconstruction captures the alternating spatial pattern of stronger and weaker antiferromagnetic correlations in all three chains. The correlators also show directly which bonds lie below the $-1/4$ threshold for pairwise entanglement. Agreement is closest for $L=4$ and $L=6$, whereas for $L=8$ the reconstructed correlations are systematically weaker, giving smaller concurrences.
These results illustrate how the first-moment formalism can image spin correlations and pairwise entanglement from experimental tunnelling spectra within the effective Heisenberg description.

\section{Other moments}\label{sec:Mhigher}
We now show that, beyond the zeroth and first moments, the moment expansion
provides access to additional spin correlations and to the
static local spin susceptibility.

\subsection{Second moment and next-neighbour correlators}
As we extend the moment analysis to higher orders, the recursive application of the Hamiltonian produces relations involving progressively longer ranged correlators. However, the complexity of these expressions increases as the moment order increases.

For the second moment, a tractable analytical expression can still be produced for the spin-1/2 Heisenberg Hamiltonian with arbitrary pairwise couplings. 
\begin{align}
    \frac{\mathcal{M}^{(2)}(i)}{g_i} &= \sum_j J_{ij}^2 \bigg( \frac{3}{8} - \frac{\langle \boldsymbol{S}(i) \cdot \boldsymbol{S}(j) \rangle}{2} \bigg) \notag \\
    &+ \sum_{j<k} J_{ji} J_{ik} \langle \boldsymbol{S}(j) \cdot \boldsymbol{S}(k) \rangle, \quad j,k \in \mathcal{N}(i) \label{eq: 2nd moment}
\end{align}
where the correlator between two sites $j,k$ which are exchange coupled to a third site $i$ appears through $\langle \boldsymbol{S}(j) \cdot \boldsymbol{S}(k) \rangle$. Unlike the first moment, this expression is only valid at zero temperature, with an additional correction appearing at finite temperatures.

This second moment, combined with the first moment, can be extremely useful to fully characterize a spin trimer in the low temperature limit, where Eq. (\ref{eq: 2nd moment}) applies. If this spin system is described by an Heisenberg Hamiltonian with all-to-all exchange couplings (whose values need not be known), then there is a total of 6 unknowns: 3 exchange couplings, plus 3 spin-spin correlators. Since both the first and the second moment depend only on these quantities via Eqs. (\ref{eq: 1st moment}) and (\ref{eq: 2nd moment}), it is possible to evaluate $\mathcal{M}^{(1)}(i)$ and $\mathcal{M}^{(2)}(i)$ from the $dI/dV$ at all sites, and solve the resulting system of non-linear equations, therefore finding all the $J_{ij}$ and $\langle \boldsymbol{S}(i) \cdot \boldsymbol{S}(j) \rangle$. Due to its non-linearity, this system may admit several solutions, so physically justified bound should constrain the final result (e.g. the sign of the exchange couplings).

A case where the system of equations can be easily inverted yielding simple analytical expressions is that of an open spin-trimer (where one of the exchange couplings is switched off). The Hamiltonian reads:
\begin{align}
    H = J_{12} \boldsymbol{S}(1)\cdot \boldsymbol{S}(2) + J_{23} \boldsymbol{S}(2)\cdot \boldsymbol{S}(3),
\end{align}
where the exchange couplings are not known. From the first moment calculation one finds $\mathcal{M}^{(1)}(1)=-J_{12} \langle \boldsymbol{S}(1) \cdot \boldsymbol{S}(2) \rangle$. Combining this with the second moment on the same site, yields the following relation:
\begin{align}
    \frac{3}{8}J_{12}^2 + \frac{\mathcal{M}^{(1)}(1)}{2} J_{12} - \mathcal{M}^{(2)}(1)=0,
\end{align}
which fully determined the exchange coupling $J_{12}$ once the quadratic equation is solved. Performing a similar analysis on the other end spin yields $J_{13}$ instead. 

Once the exchange couplings are determined, all spin-spin correlators follow directly from the first and second moments according to the following relations:
\begin{align}
    \langle \boldsymbol{S}(1) \cdot \boldsymbol{S}(2)\rangle &= - \frac{\mathcal{M}^{(1)} (1)}{J_{12}}, \label{eq:Jtrimer1}\\
    \langle \boldsymbol{S}(2) \cdot \boldsymbol{S}(3)\rangle &= - \frac{\mathcal{M}^{(1)} (3)}{J_{23}} \label{eq:Jtrimer2}\\
    \langle \boldsymbol{S}(1) \cdot \boldsymbol{S}(3)\rangle &=  \frac{\mathcal{M}^{(2)} (2) - \mathcal{M}^{(2)} (1) - \mathcal{M}^{(2)} (3)}{J_{12} J_{23}},
    \label{eq:Jtrimer3}
\end{align}
where even the correlator between sites 1 and 3, which are not connected via exchange couplings, can be reconstructed. Eqs.(~\ref{eq:Jtrimer1}), (\ref{eq:Jtrimer2}) and (\ref{eq:Jtrimer3}) provide an explicit example of the one-to-one functional relation  between spin couplings and correlators ~\cite{murta22,quintanilla22}.


\subsection{Inverse moment $(n=-1)$and local susceptibility}
\label{sec:inverse_moment}

The $n=-1$ moment is known to relate to the static susceptibility in systems with spin translational invariance\cite{hohenberg74}. Here we show the relation between the inverse moment $\mathcal{M}^{(-1)}(i)$ and the static local spin susceptibility, for arbitrary spin systems with a unique ground state,  $|G\rangle$, separated from all excited states by a finite gap $\Delta \gg k_B T$. In this limit, we have:
\begin{equation}
\frac{\mathcal{M}^{(-1)}(i)}{g_i} = \sum_{n\neq G}\sum_{a=x,y,z} \frac{|\langle n|S_a(i)|G\rangle|^2}{E_n-E_G}.
\label{eq:inverse_moment_spectral}
\end{equation}
We note that, at $T=0$, the $d^2I/dV^2$ vanishes at small $V$ and therefore there is no divergence at $V=0$. At small $T$ it can be shown that $d^2I/dV^2$ is linear in $V$ for $V \ll \Delta$ so that, again, there is no divergence of the integral near zero bias.

The result in Eq. (\ref{eq:inverse_moment_spectral})  can be related to the local susceptibility, defined   as the response to a local perturbation $\delta H=-\sum_{b=x,y,z} h_b(i)S_b(i)$:
\begin{equation}
\chi_{ab}(i) = \left.\frac{\partial\langle S_a(i)\rangle} {\partial h_b(i)} \right|_{\boldsymbol{h}(i)=0},
\end{equation}
where $h_b(i)$ has units of energy. First-order perturbation theory gives\cite{kubo57}
\begin{equation}
\chi_{ab}(i) = 2\,\mathrm{Re}\sum_{n\neq G} \frac{\langle G|S_a(i)|n\rangle \langle n|S_b(i)|G\rangle }{E_n-E_G}.
\end{equation}
Taking the trace over the three spin components therefore yields
\begin{equation}
\mathrm{Tr}\,\boldsymbol{\chi}(i) = \frac{2\,\mathcal{M}^{(-1)}(i)}{e g_i}.
\label{eq:inverse_moment_susceptibility}
\end{equation}
We note that this relation does not require spin-rotational invariance and is valid for arbitrary  spin.  For instance, it could  be used to determine  the ground state spin susceptibility of  single integer spins where single ion anisotropy results in a unique ground state\cite{hirjibehedin06,tsukahara09}, as well as Haldane spin chains even in the presence of non-linear exchange\cite{Mishra21}.

\section{Genuine tripartite entanglement \label{sec:GTE}}


We now consider the situation of three exchange-coupled $S=1/2$ spins. This situation arises naturally in clusters of chemisorbed atomic hydrogen in graphene\cite{vina26}, and also in $L=3$ nanographene spin chains made with $S=1/2$ radicals, such as phenalenyl\cite{yuan25}, Clar's goblets\cite{zhao24} and olimpecenes\cite{zhao25}. 

Three interacting spins can support pairwise entanglement, as in a singlet dimer with a separable third spin, or genuine tripartite entanglement, in which no spin can be separated from the other two. These forms of entanglement are not mutually exclusive: pairwise correlations can coexist with genuine tripartite entanglement. Their relative importance depends on the signs and strengths of the exchange interactions. The concurrence fill, discussed in Eq.~(\ref{eq:concurrence_fill}) of Sec. \ref{sec:entanglement} allows us to follow how entanglement shared by all three spins develops as the couplings evolve from a strongly dimerized configuration towards a collectively coupled trimer.


\begin{figure}[t]
\centering
\includegraphics[width=0.45\textwidth]{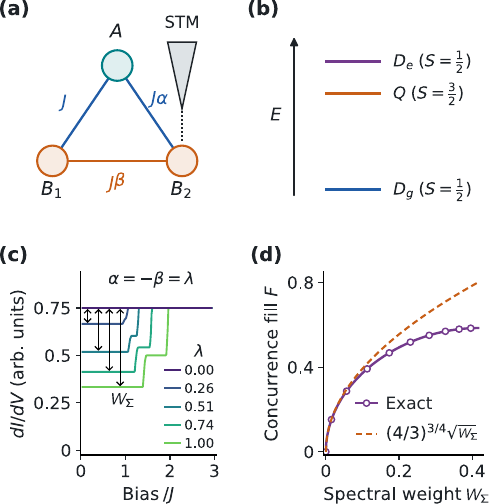}
\caption{\textbf{Genuine tripartite entanglement from collective spin excitations.}
(a) Schematic representation of the trimer, and its exchange couplings.
(b) The two $S=1/2$ doublets and the $S=3/2$ quartet of the Heisenberg trimer; energies and splittings are schematic.
(c) Simulated local IETS at $B_2$ for $\alpha=-\beta=\lambda=0,0.1,0.25,0.5,1$. 
(d) Ground-state concurrence fill versus the combined exchange weight $W_\Sigma = W_D + W_Q$. The solid curve is the exact relation and the dashed curve its weak-coupling square-root limit. 
Parameters: $J=10\,\mathrm{meV}$ and $T=300\,\mathrm{mK}$. Spectra include thermal occupations and thermal electron broadening.
}
\label{fig:trimer_quartet}
\end{figure}

We consider the following Hamiltonian for three spins $A,B_1$ and $B_2$, with antiferromagnetic coupling between spins $A$ and $B_{1/2}$, and ferromagnetic coupling between spins $B_1$ and $B_2$:
\begin{align}
    H &=J\big[\boldsymbol{S}(A)\cdot\boldsymbol{S}(B_1) +\alpha\boldsymbol{S}(A)\cdot\boldsymbol{S}(B_2) +\beta\boldsymbol{S}(B_1)\cdot\boldsymbol{S}(B_2) \big]
    \label{eq:trimer_hamiltonian}
\end{align}
where $J>0$, $\alpha\geq0$, $\beta\leq0$. The exchange spectrum consists of two $S=1/2$ doublets, $D_g$ and $D_e$, and one $S=3/2$ quartet, $Q$. The two antiferromagnetic $A$--$B$ couplings and the ferromagnetic $B_1$--$B_2$ coupling favor a doublet ground multiplet.  Assuming negligible occupation of the excited multiplets, IETS therefore exhibits at most two inelastic thresholds, at the exchange gaps $\Delta_D$ and $\Delta_Q$ from $D_g$ to $D_e$ and $Q$, respectively. The spin system and its energy levels are depicted in Fig. \ref{fig:trimer_quartet}.

We describe both $S=1/2$ doublets ($D_g$ and $D_e$) by coupling the singlet, $|S\rangle$, and triplet, $|T,M = 0,1\pm 1\rangle$, of the $A$--$B_1$ pair to spin $B_2$:
\begin{equation}
\begin{aligned}
|s,+\rangle&=|S\rangle|\uparrow\rangle,\quad
|s,-\rangle=|S\rangle|\downarrow\rangle,\\
|t,+\rangle&=
\sqrt{\frac23}|T,+1\rangle|\downarrow\rangle
-\sqrt{\frac13}|T,0\rangle|\uparrow\rangle,\\
|t,-\rangle&=
\sqrt{\frac13}|T,0\rangle|\downarrow\rangle
-\sqrt{\frac23}|T,-1\rangle|\uparrow\rangle.
\end{aligned}
\label{eq:trimer_doublet_basis}
\end{equation}
The eigenstates of Eq. (\ref{eq:trimer_hamiltonian}) are then written as the orthogonal combinations:
\begin{equation}
\begin{aligned}
|D_g,\pm\rangle&=a|s,\pm\rangle+b|t,\pm\rangle,\\
|D_e,\pm\rangle&=-b|s,\pm\rangle+a|t,\pm\rangle,
\end{aligned}
\label{eq:trimer_doublets}
\end{equation}

where $a,b$ are real, obey $a^2+b^2=1$ and depend on the exchange couplings. Note that the state  $|s,\pm\rangle$  is  biseparable, with the $B_2$ spin decoupled from the $A,B_1$ pair. Therefore, one can already anticipate that tripartite entanglement in the ground state requires a finite value of $b$.

The quartet wave functions are independent of the exchange parameters, since three spins $1/2$ contain a single fully symmetric $S=3/2$ multiplet. The states $|D_g,+\rangle$ and $|D_g,-\rangle$ form a doubly degenerate $S=1/2$ ground state. As discussed in Sec. \ref{sec:entanglement}, for the case of a spin-rotational invariant $S=1/2$ degenerate ground state, the concurrence fill can be determined from each of the $S_z$ components. In what follows, we will focus our analysis on the $|G\rangle = |D_g,+\rangle$ state.

We now make two key observations. First, when $b=0$, a tip probing spin $B_2$ cannot induce inelastic transitions from the ground-state doublet to either the excited doublet or the quartet. Transitions between the two doublets are forbidden because their $A$--$B_1$ spin wave functions are orthogonal, $\langle s|t\rangle=0$. Likewise, transitions to the $S=3/2$ quartet are forbidden because a singlet $A$--$B_1$ pair coupled to spin $B_2$ can only form a state with total spin $S=1/2$. Consequently, these transitions acquire finite spectral weight only when $b\neq 0$. Second, as pointed out before, at $b=0$, the ground state is biseparable, with the $A$--$B_1$ singlet disentangled from spin $B_2$. Together, these observations suggest a direct connection between genuine tripartite entanglement and the inelastic spectral weights measured at $B_2$, which we now derive.

In order to compute the concurrence fill we use Eq. (\ref{eq:concurrence_fill}) and the relation of the one-tangles with the local magnetization of Eq. (\ref{eq:one_tangle}).  From Eq. (\ref{eq:trimer_doublets}), the magnetizations read:
\begin{equation}
m_A=\frac{b^2}{3}-\frac{ab}{\sqrt3},\;
m_{B_1}=\frac{b^2}{3}+\frac{ab}{\sqrt3},\;
m_{B_2}=\frac12-\frac{2b^2}{3}
\label{eq:trimer_magnetizations}
\end{equation}
where $m_i=\langle G|S_z(i)|G\rangle$ and the transverse averages vanish; from $m_{B_2}$ one establishes $0\leq b^2\leq3/4$. The concurrence fill consequently reads
\begin{equation}
F_{AB_1B_2}=
\left(\frac{256}{27}\right)^{1/4}
|b|\left|1-\frac43b^2\right|.
\label{eq:trimer_fill_b}
\end{equation}

For each excited multiplet $\mathcal{X}=D_e,Q$, we define its summed local transition weight
\begin{equation}
W_{\mathcal{X}}\equiv
\sum_{X \in \mathcal{X}}\sum_{\gamma=x,y,z}
|\langle X|S_\gamma(B_2) |G\rangle|^2.
\end{equation}
Direct evaluation gives
\begin{equation}
W_D=\frac43b^2(1-b^2),\qquad
W_Q=\frac23b^2.
\label{eq:trimer_multiplet_weights}
\end{equation}
Hence, it becomes self-evident that concurrence fill for the ground state and the weight of the inelastic transitions out of the doublet space are all controlled by $b$.
Both exchange contributions can be combined:
\begin{equation}
W_\Sigma=W_D+W_Q
=2b^2-\frac43b^4
\label{eq:trimer_combined_weight}
\end{equation}
Solving for $b^2$ and defining $r_\Sigma=\sqrt{1-4W_\Sigma/3}$, yields $b^2=3(1-r_\Sigma)/4$ and hence we can derive the relation between the concurrence fill for the ground state and the inelastic weights out of the ground state doublet sector: 
\begin{equation}
F_{AB_1B_2}=
\left(\frac{16}{3}\right)^{1/4}
r_\Sigma\sqrt{1-r_\Sigma}.
\label{eq:trimer_fill_combined}
\end{equation}
%
%
The combined conductance increment  does not require the transitions to the excited doublet and quartet to be resolved spectrally, which can be convenient when their splitting is smaller than the experimental resolution.  In the weak-coupling limit (small $\alpha,\beta$), these relations reduce to
\begin{equation}
\begin{split}
F_{AB_1B_2}\simeq\left(\frac43\right)^{3/4}\sqrt{W_\Sigma}.
\end{split}
\label{eq:trimer_fill_weak}
\end{equation}

Figure~\ref{fig:trimer_quartet} illustrates these results along $\alpha=-\beta=\lambda$, $0\leq\lambda\leq1$. For this particular parameter choise, one finds $b^2=[1-(1+3\lambda^2)^{-1/2}]/2$. 
The appearance of exchange steps at $B_2$ accompanies the growth of genuine tripartite entanglement. Along this ramp, $W_D/W_Q$ decreases from $2$ to $3/2$, so that the doublet excitation is stronger than the quartet excitation (see Fig.~\ref{fig:trimer_quartet}c).  This intensity ordering can help assign the experimental thresholds. In Fig. \ref{fig:trimer_quartet}d we show the exact relation between the concurrence fill and the combined exchange weight, and its square-root limit. At $\lambda=1$, $W_\Sigma=5/12$ and $F=4/3^{7/4}$.

\begin{figure*}[t]
\centering
\includegraphics[width=0.95\textwidth]{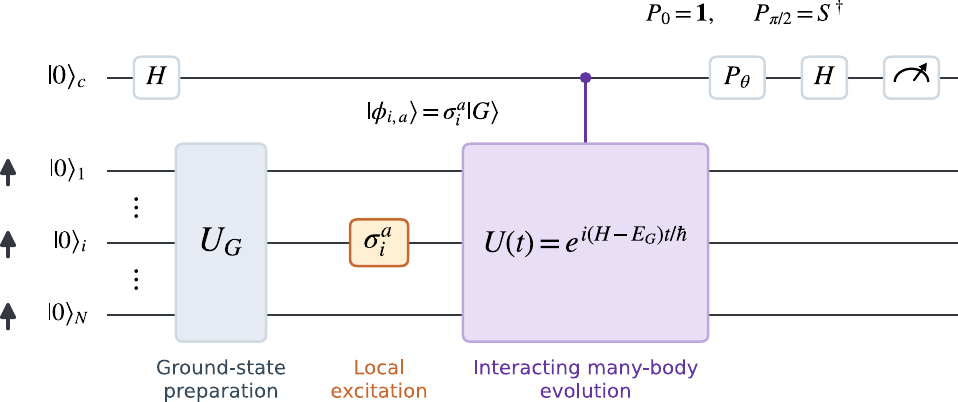}
\caption{
\textbf{Digital quantum simulation of local spin IETS using a Hadamard test.}
The preparation unitary $U_G$ initializes the spin register
in its ground state, $U_G|0\rangle^{\otimes N}=|G\rangle$.
A single-qubit Pauli gate creates the normalized state
$|\phi_{i,a}\rangle=\sigma_a(i)|G\rangle=2S_a(i)|G\rangle$, simulating the tunneling exchange interaction on the spin under the  STM tip.
The ancillary qubit $c$ controls evolution under the
interacting Hamiltonian,
$U(t)=e^{i(H-E_G)t/\hbar}$.
With $P_0=\boldsymbol{1}$ or $P_{\pi/2}=S^\dagger$,
the final ancilla measurement gives the real or imaginary
part of $\mathcal{L}_{i,a}(t)$, respectively.
Fourier transformation over the evolution time recovers
the excitation energies and local spin spectral weights.
The collective evolution explicitly distinguishes the
protocol from a measurement confined to one spin.
Here $H$ on the ancilla wire denotes a Hadamard gate,
while $H$ in $U(t)$ denotes the spin Hamiltonian.
}
\label{fig:iets_hadamard}
\end{figure*}

\section{A quantum-computational analogue of IETS}
\label{sec:quantum_circuit}

At first sight, extracting entanglement from STM-based IETS may appear surprising: how can a local probe reveal correlations between different spins? The excitation induced by the tunneling
electron is local, but the measured response is governed by the interacting many-body Hamiltonian. The exchange interactions remain active throughout the experiment, and the excitation
energies and transition weights therefore encode properties of the entire spin system. Energy-resolved local spectroscopy therefore provides information beyond that which is contained in the
single-site reduced density matrix.

Here we introduce a digital quantum analogue of STM-based IETS that makes explicit the multi-qubit operations underlying its access to many-body correlations. Our construction, depicted in Fig. \ref{fig:iets_hadamard}, builds on quantum algorithms for dynamical response functions based on phase estimation~\cite{kosugi20,sels21} and interferometric spectroscopy~\cite{zhai26}, alongside
related quantum simulations of linear response~\cite{kokcu24}. 

We consider $N$ interacting spin-$1/2$ degrees of freedom, represented by $N$ qubits. Ground-state preparation can be approached through variational optimization~\cite{murta24}, quantum imaginary-time evolution~\cite{motta20}, or, for suitable spin models, dedicated valence-bond-solid circuits~\cite{murta23} and adiabatic tensor-network protocols~\cite{wei23}. The spectroscopic procedure below is independent of the chosen preparation method.

The local action of the tunneling electron on site $i$ is represented by one of the three spin operators $S_a(i)$, with $a=x,y,z$. Since $S_a(i)=\sigma_a(i)/2$, the normalized state following this operation is obtained by applying a single-qubit Pauli gate,
\begin{equation}
    |\phi_{a}(i)\rangle =\sigma_a(i)|G\rangle =2S_a(i)|G\rangle.
    \label{eq:local_excitation_state}
\end{equation}
with $|G\rangle$ represent once again the ground state.

The information probed by IETS is contained in the energy distribution of this locally excited state. To access it interferometrically~\cite{somma02}, we introduce the unitary
\begin{equation}
    U(t)=e^{i(H-E_G)t/\hbar}
\end{equation}
and evaluate the return amplitude
\begin{align}
    \mathcal{L}_{a}(i,t) &=\langle\phi_{a}(i)|U(t)|\phi_{a}(i)\rangle \notag \\
    &=4\sum_X |\langle X|S_a(i)|G\rangle|^2 e^{i(E_X-E_G)t/\hbar}.  
    \label{eq:iets_return_amplitude}
\end{align}
A Hadamard test\cite{somma02} measures its real and imaginary parts through the interference of an ancillary qubit controlling the many-body evolution. Repeating the measurement for different
evolution times and Fourier transforming gives
\begin{eqnarray}
    \frac{1}{2\pi\hbar} \int_{-\infty}^{\infty}dt\, e^{-i\varepsilon t/\hbar}\mathcal{L}_{a}(i,t) =\nonumber \\
    4\sum_X|\langle X|S_a(i)|G\rangle|^2 \delta\!\left[\varepsilon-(E_X-E_G)\right].
\label{eq:quantum_iets_spectrum}
\end{eqnarray}
Thus, the peak positions yield the excitation energies, while their weights yield the spin transition probabilities. Summing the three channels reproduces the local spin spectral weight entering the IETS conductance. Finite evolution times and discrete sampling set the spectral resolution and accessible energy range.

Alternatively, quantum phase estimation applied to $|\phi_{a}(i)\rangle$ samples the eigenenergies $E_X-E_G$ with probabilities $4|\langle X|S_a(i)|G\rangle|^2$, providing the same spectral information through repeated measurements\cite{cruz20}. 

Both constructions separate a local preparation step from a collective spectral readout. Although $\sigma_a(i)$ acts on one qubit, evolution under the interacting Hamiltonian generally cannot be factorized into independent single-qubit unitaries: its digital implementation requires entangling gates between spins. The experimental counterpart is the persistent exchange interaction that determines the energy-resolved response to electron scattering. The extraction of entanglement from IETS therefore relies on local excitation combined with interacting many-body dynamics and spectral resolution. This correspondence connects STM spin spectroscopy with quantum algorithms for measuring many-body correlation functions and spectra.

\section{Discussion and Conclusions \label{sec:disc}}

\subsection{Summary of main results }
Traditionally, STM-based IETS  of on-surface spins and spin lattices has been used  to  infer spin excitation energies and spin-matrix elements. The main theme of this work is to show that STM-based IETS has much more to offer. We have introduced a key tool of {\em information compression}, the moments $ \mathcal{M}^{(n)}(i)$ in Eq. (\ref{eq: moment def}).   They can be exploited in several ways:
\begin{itemize}
    \item For $n=0$,  and non-degenerate ground state $|G\rangle$ with negligible thermal depletion, the moment $ \mathcal{M}^{(0)}(i)$, that measures the height of all inelatic steps for site $i$, provides the variance of the spin   at that site.  We note that the variance is minimal for eigenstates of $S_a(i)$. Therefore, for integer spins with quantum spin tunneling this predicts  that inelastic signal will be maximal at zero applied field and it will shrink as a field is applied, consistent with experimental observations for $S=2$ Fe on  Cu$_2$N \cite{hirjibehedin07} and $S=1$ Fe phtalocyanines \cite{tsukahara09}.
    \item The $n=0$  moment distinguishes antiferromagnetic from ferromagnetic spin-$1/2$ Heisenberg dimers. At temperatures well below the exchange gap and zero magnetic field, the normalized inelastic weight is $\mathcal{M}^{(0)}(i)/g_i=3/4$ and $1/4$ for a singlet and a triplet ground state, respectively.  Figure~\ref{fig:AA_AB_step_height} shows qualitative agreement with experimental spectra of hydrogen dimers on graphene.
    \item In the specific case of  $S=1/2$,  for a unique ground state without thermal depletion, $ \mathcal{M}^{(0)}(i)$ measures the one-tangle for each spin, that relates to the length of the average local magnetization, c.f. Eq. (\ref{eq:one_tangle}).   Hence STM-based IETS can be used to image the  non-trivial local magnetization/one-tangle maps of antiferromagnetic spin chains with a local Zeeman perturbation\cite{delCastillo26} that  have a unique ground state.  
    \item The $n=1$ moment,  $ \mathcal{M}^{(1)}(i)$, gives the spatial distribution of exchange energy, which then yields the ground state energy (see Eq. (\ref{eq: 1st moment bilinear})). For spin-chains with spin-rotational invariance, it can be used to map the Wooters pairwise entanglement along the chain.  We illustrate the potential of  the method with raw experimental data \cite{zhao25}, obtaining experimental images of entanglement taken with STM.
    \item The $n=2$  moment, $ \mathcal{M}^{(2)}(i)$, provides correlations between spins connected through a common neighbour  in the  spin-$1/2$ Heisenberg systems with first neighbor exchange, extending the information accessible from the first moment (see Eq. (\ref{eq: 2nd moment})).
     \item The $n=-1$ moment, determines the trace of the static local spin susceptibility for a unique gapped ground state with negligible thermal depletion (see  Eq.~\eqref{eq:inverse_moment_susceptibility}). This results holds  for arbitrary spin and arbitrary spin Hamiltonians
\end{itemize}

We have also established a connection between STM-based IETS and the concurrence fill, a faithful indicator of genuine tripartite entanglement for pure three-spin states (see Eq. (\ref{eq:trimer_fill_combined})) . For exchange-coupled spin-$1/2$ trimers, this connection allows us to infer how entanglement shared by all three spins develops from the spectral weight of collective spin excitations.

Table~\ref{tab:iets_entanglement} summarizes the connections established here between STM-based IETS and entanglement in on-surface spin systems. 

At last, we introduced a digital quantum circuit that reproduces the spectral information accessed by spin IETS. This construction makes explicit how a local probe accesses many-body entanglement: although the initial spin operation acts on a single qubit, the subsequent evolution under the interacting Hamiltonian involves multi-qubit operations that encode nonlocal correlations in the measured spectrum.

\bluemark{\begin{table*}[t]
\centering
\small
\renewcommand{\arraystretch}{1.4}
\begin{tabular}{@{}lllll@{}}
\toprule
IETS information & Information obtained
& System & Conditions & Where \\
\midrule

$N_{\mathrm{steps}}>2$
& Entanglement witness
& Spin-$1/2$ dimer
& $p_G\simeq1$; local probe
& Sec.~\ref{sec:IETS}
\\

$\mathcal{M}^{(0)}(i)$
& One-tangle $\tau_i$
& $N$ spins $1/2$
& Unique GS; $k_{\mathrm B}T\ll\Delta$
& Eq.~\eqref{eq:main1}
\\

$\mathcal{M}^{(1)}(i)$
& Concurrence $C_{ij}$
& Spin-$1/2$ Heisenberg tree
& SU(2)-invariant state; known $J_{ij}$
& Eqs.
\eqref{eq:concurrence_correlator}, ~\eqref{eq: 1st moment bilinear}
\\

$\mathcal{M}^{(2)}(i)$
& $\langle\boldsymbol{S}(j)\cdot\boldsymbol{S}(k)\rangle$,
$j,k\in\mathcal{N}(i)$
& Spin-$1/2$ Heisenberg system
& $T=0$; weighted combinations
& Eq.~\eqref{eq: 2nd moment}
\\

$\mathcal{M}^{(-1)}(i)$
& Local  susceptibility $\mathrm{Tr}\,\boldsymbol{\chi}(i)$
& Arbitrary $S_i$ and Hamiltonian
& Unique GS; $k_{\mathrm B}T\ll\Delta$
& Eq.~\eqref{eq:inverse_moment_susceptibility}
\\

$W_D+W_Q$
& Concurrence fill $F_{AB_1B_2}$
& Spin-$1/2$ Heisenberg trimer
& Ground doublet only; specified $J_{ij}$ signs
& Eq.~\eqref{eq:trimer_fill_combined}
\\

\bottomrule
\end{tabular}
\caption{
Connections between STM-based IETS and spin entanglement, 
spin correlators and local spin susceptibility.
All relations assume perturbative spin-assisted tunnelling.
$N_{\mathrm{steps}}$ counts resolved inelastic thresholds
from a single initial state; its witness certifies
entanglement in the eigenbasis, not necessarily in the GS.
$\Delta$ is the excitation gap, $\rho_T$ the thermal state,
and $W_D,W_Q$ the normalized weights into the excited
doublet and quartet.
}
\label{tab:iets_entanglement}
\end{table*}}

\subsection{Approximations and limitations}

Our theory neglects entanglement between the surface spins and the substrate electrons. This limits the identification of local spin fluctuations with entanglement within the spin system: the one-tangle inferred assuming a pure state becomes an upper bound when the spin system is entangled with its environment~\cite{rungta01,coffman00}. This approximation is more likely to fail for degenerate ground states or systems with a small excitation gap, where substrate coupling can compete with the intrinsic spin interactions. Decoupling layers can reduce this coupling.

We also neglect third-order exchange corrections, which produce overshoots at the inelastic thresholds and bias-dependent conductance above the steps~\cite{ternes15}. We have assessed their impact, through numerical calculation (not shown), and found that they can significantly modify the spectral moments, albeit in a controlled manner. Substrate and phonon contributions can also change the lineshapes. 

There are two ways around these limitations. First, introducing decoupling layers between on-surface spins and the conducting substrate reduces the Kondo coupling, and can result in very flat $dI/dV$ lineshapes, in agreement with the second order perturbation theory\cite{huang25}. Alternatively, we have verified that fitting the experimental $d^2I/dV^2$ with Lorentzian peaks
and integrating the fitted curves over the experimental bias window gives better agreement with our theoretical predictions (not shown). This empirical procedure reduces the sensitivity to lineshape distortions. 

\subsection{Conclusions}

We have established several connections between STM-based IETS and spin entanglement, providing ways to detect, quantify and image entanglement in on-surface spin systems. 

In particular, our analysis of experimental nanographene-chain spectra illustrates how IETS can resolve the spatial distribution of pairwise concurrence (see Fig.~\ref{fig:exp_vs_theory_Chenxiao}). The relation between the zeroth moment and the one-tangle also opens a route to imaging entanglement between individual spins and the rest of a spin lattice. A promising application is a topological spin chain subject to a magnetic field acting on a single spin, where the spatial redistribution of magnetization would provide a corresponding image of the one-tangle in the pure-ground-state regime.

We have also proposed a quantum circuit that reproduces the spectral information accessed by STM-based IETS (Fig.~\ref{fig:iets_hadamard}). This construction makes explicit how local spectroscopy accesses many-body entanglement through the interacting spin dynamics.

Together, these results establish a bridge between atomic-scale scanning tunnelling spectroscopy and quantum information. They give familiar spectroscopic observables a new role as quantitative probes of entanglement, extending the capabilities of STM towards imaging quantum correlations. This connection opens new ways to explore quantum magnetism in engineered spin systems, with spatially resolved entanglement as an experimentally accessible property.

\vspace{0.5cm}
\textit{Note added.---} During the final stages of this work,
we became aware of a related preprint addressing the
certification of nonlocal spin entanglement from local
tunneling spectra~\cite{kazemikhah26}.

\section*{Acknowledgments}

We acknoweledge Roman Fasel,   Oliver Gr\~onig, Pascal Ruffieux and Gonçalo Catarina for fruitful discussions.
J.F.R.   and J.C.H. acknowledge financial support from 
SNF Sinergia (Grant Pimag), and
the QUANTUM\_IBER\_IA project
(0590\_QUANTUM\_IBER\_IA\_6\_E), co-funded by the European
Union through the European Regional Development Fund
under the Interreg VI-A Spain--Portugal (POCTEP)
2021--2027 programme.
%
I. B  and B.V.B.   acknowledge financial support from the Spanish Ministry of Science and Innovation, through project PID2023-149106NB-I00,  the María de Maeztu Program for Units of Excellence in R \& D (grant no. CEX2023–001316-M), the Comunidad de Madrid and the Spanish State through the Recovery, Transformation and Resilience Plan [Materiales Disruptivos Bidimensionales (2D), (MAD2DCM)-UAM Materiales Avanzados] the European Union through the Next Generation EU funds and and by “Artificial Quantum Matter: From 2D Materials to Spin Lattice Systems” BBVA Foundation Fundamentos Program 2024. B.V.-B. acknowledges funding from the Spanish Ministerio de Universidades through the PhD grant FPU2022/03675.
We acknowledge the use of ChatGPT (OpenAI) for assistance with
manuscript editing, literature searches, and the preparation of
figures and computational scripts. 
 ChatGPT also assisted in
deriving the relation between the first spectral moment and
the one-tangle, and in assessing the validity of the
first-moment expression at finite temperature. All content
was reviewed and verified by the authors, who take full
responsibility for the final manuscript.

\appendix

\section{Derivation of the first-moment sum rule}
\label{app:first_moment}

We derive the first-moment sum rule, including its
validity at finite temperature. With the definition
of $\mathcal{M}^{(1)}(i)$ in
Eq.~\eqref{eq: moment def}, and keeping the tunneling
prefactor explicit, the result is
\begin{equation}
\frac{e\,\mathcal{M}^{(1)}(i)}{g_i}
=
\frac{1}{2}\sum_m p_m\sum_{a=x,y,z}
\langle m|[S_a(i),[H,S_a(i)]]|m\rangle .
\label{eq:app_first_moment}
\end{equation}

\subsection{First moment of the thermal kernel}

We first consider a single transition with energy
$\Delta$, which can be positive or negative.
Introducing the energy variable $v=eV$, we write
the kernel of Eq.~\eqref{eq:IETSkernel} as
\begin{equation}
{\cal K}_T(v/e,\Delta)
=f_T'(v-\Delta)+f_T'(-v-\Delta),
\end{equation}
where primes denote derivatives with respect to
the argument and
\begin{equation}
f_T(x)=\frac{x}{1-e^{-\beta x}},
\qquad \beta=(k_BT)^{-1}.
\end{equation}
The identity
\begin{equation}
f_T(-x)=f_T(x)-x
\end{equation}
implies that $h_T(x)\equiv f_T''(x)$ is even.
Its integral is
\begin{equation}
\int_{-\infty}^{\infty}h_T(x)\,dx
=f_T'(\infty)-f_T'(-\infty)=1,
\label{eq:app_h_normalization}
\end{equation}
since $f_T'(\infty)=1$ and $f_T'(-\infty)=0$.

Differentiating the kernel and using the evenness
of $h_T$ gives
\begin{align}
\frac{\partial}{\partial v}
{\cal K}_T(v/e,\Delta)
&=h_T(v-\Delta)-h_T(-v-\Delta)
\nonumber\\
&=h_T(v-\Delta)-h_T(v+\Delta).
\end{align}
Its first moment is therefore
\begin{align}
J_T(\Delta)
&\equiv\int_0^\infty dv\,v
\frac{\partial}{\partial v}
{\cal K}_T(v/e,\Delta)
\nonumber\\
&=\int_0^\infty dv\,v\,h_T(v-\Delta)
-\int_0^\infty dv\,v\,h_T(v+\Delta).
\label{eq:app_kernel_integral}
\end{align}
In the second integral, the substitution $u=-v$
gives
\begin{align}
\int_0^\infty dv\,v\,h_T(v+\Delta)
&=-\int_{-\infty}^{0}du\,u\,h_T(\Delta-u)
\nonumber\\
&=-\int_{-\infty}^{0}du\,u\,h_T(u-\Delta),
\end{align}
where the last equality again uses evenness.
Consequently, Eq.~\eqref{eq:app_kernel_integral}
becomes a full-line integral:
\begin{align}
J_T(\Delta)
&=\int_{-\infty}^{\infty}dv\,v\,h_T(v-\Delta)
\nonumber\\
&=\int_{-\infty}^{\infty}dx\,(x+\Delta)h_T(x)
\nonumber\\
&=\Delta .
\label{eq:app_kernel_moment}
\end{align}
Here we used $x=v-\Delta$, the vanishing integral
of the odd function $xh_T(x)$, and
Eq.~\eqref{eq:app_h_normalization}.
This result holds for either sign of $\Delta$
and at any temperature.

\subsection{From transition weights to the double commutator}

The spin-independent conductance background does not
contribute to the derivative. Inserting
Eq.~\eqref{eq:GIETS} into the moment definition gives
\begin{equation}
\mathcal{M}^{(1)}(i)
=g_i\sum_{m,n}p_m W_{mn}^{(i)}
\int_0^\infty dV\,V
\frac{\partial{\cal K}_T(V,E_n-E_m)}{\partial V}.
\end{equation}
Changing variables to $v=eV$ and applying
Eq.~\eqref{eq:app_kernel_moment}, we find
\begin{equation}
\frac{e\,\mathcal{M}^{(1)}(i)}{g_i}
=\sum_{m,n}p_m(E_n-E_m)
\sum_{a=x,y,z}
|\langle n|S_a(i)|m\rangle|^2 .
\label{eq:app_spectral_sum}
\end{equation}
Thus, thermal broadening preserves the
energy-weighted spectral sum.

For each initial eigenstate $|m\rangle$ and spin
component $a$, completeness gives
\begin{align}
&\sum_n(E_n-E_m)
|\langle n|S_a(i)|m\rangle|^2
\nonumber\\
&\quad =
\sum_n
\langle m|S_a(i)|n\rangle
(E_n-E_m)
\langle n|S_a(i)|m\rangle
\nonumber\\
&\quad =
\langle m|S_a(i)(H-E_m)S_a(i)|m\rangle .
\label{eq:app_completeness}
\end{align}
To express this result as a double commutator,
we expand
\begin{equation}
[S_a(i),[H,S_a(i)]]
=
2S_a(i)HS_a(i)-S_a^2(i)H-HS_a^2(i).
\end{equation}
Since $H|m\rangle=E_m|m\rangle$, its diagonal
matrix element is
\begin{align}
&\langle m|[S_a(i),[H,S_a(i)]]|m\rangle
\nonumber\\
&\quad =
2\langle m|S_a(i)HS_a(i)|m\rangle
-2E_m\langle m|S_a^2(i)|m\rangle
\nonumber\\
&\quad =
2\langle m|S_a(i)(H-E_m)S_a(i)|m\rangle .
\end{align}
Combining this identity with
Eqs.~\eqref{eq:app_spectral_sum} and
\eqref{eq:app_completeness} proves
Eq.~\eqref{eq:app_first_moment}.

The result applies to arbitrary local spins and
ground-state degeneracy, within the equilibrium
perturbative tunneling theory used here. Its
finite-temperature validity requires retaining both
terms of the thermal kernel and all thermally
occupied initial states. No pure-ground-state
approximation is needed.


%

\end{document}